\RequirePackage{silence}
\documentclass[aps,reprint,
dvipsnames]{revtex4-1}
\pdfoutput=1
\usepackage{blindtext}

\usepackage{graphicx}% Include figure files
\usepackage{dcolumn}% Align table columns on decimal point
\usepackage{bm}% bold 
\usepackage{amssymb}
\usepackage{color}
\usepackage[T1]{fontenc} 
\usepackage{mathrsfs,amsfonts,dsfont}
\usepackage{nccmath}
\usepackage{amstext}
\usepackage{mathtools}
\usepackage{tablefootnote}
\usepackage[inline]{enumitem}
\graphicspath{{graphics/}}
\usepackage[english]{babel}

\usepackage[table]{xcolor}

\usepackage{wrapfig}
\usepackage{lipsum} 
\usepackage{blindtext}

\usepackage{braket}
\usepackage{bbm} 
\usepackage[normalem]{ulem}

\usepackage[nolist,withpage]{acronym}
\makeatletter
\AtBeginDocument{%
  \renewcommand*{\AC@hyperlink}[2]{%
    \begingroup
      \hypersetup{hidelinks}%
      \hyperlink{#1}{#2}%
    \endgroup
  }%
}
\makeatother

\usepackage[colorlinks=true,citecolor=blue,linkcolor=blue]
{hyperref}

\usepackage{amsthm}

\newtheorem{result}{Result}

\definecolor{martin}{rgb}{0,.4,1}

\DeclareMathOperator{\Tr}{Tr}
\DeclareMathOperator{\Ddiamond}{D_{\diamond}}

\DeclareMathOperator{\Idiamond}{I_{\diamond}}

\newcommand\mdoubleplus{\mathbin{+\mkern-10mu+}}

 \hypersetup{pdftitle = {Distributed quantum incompatibility},
       pdfauthor = {Lucas Tendick, Hermann Kampermann, Dagmar Bruß},
       pdfsubject = {Quantum information theory}, 
       pdfkeywords = {quantum, assemblage, resource, theory, 
                      SDP, semidefinite program, incompatibility, mutually unbiased bases, MUB, steering, EPR, nonlocality, Bell,
                      entanglement, quantification, distance, metric,
                      monotone, quantifier assemblage, prepare-and-measure, 
                      distance between quantum measurements, 
                      jointly measurable, Bell inequality, %Bell inequality, 
                      quantum advantage, convex, optimization, dual,
                      gain, incompatibility gain, triangle inequality,
                      measurement splitting, parent POVM, polygamie,
                      polygamous, genuine, genuine triplewise, pairwise incompatible, structures, splitting, bound, decomposition,
              }
      }

\begin{document}
\title{Distributed quantum incompatibility}
\author{Lucas Tendick}
\email{lucas.tendick@hhu.de}
\author{Hermann Kampermann}
\author{Dagmar Bru\ss}
\affiliation{Institute for Theoretical Physics, Heinrich Heine University D\"usseldorf, 
D-40225 D\"usseldorf, Germany}

\begin{abstract}
Incompatible, i.e. non-jointly measurable quantum measurements are a necessary resource for many information processing tasks. It is known that increasing the number of distinct measurements usually enhances the incompatibility of a measurement scheme. However, it is generally unclear how large this enhancement is and on what it depends. Here, we show that the incompatibility which is gained via additional measurements is upper and lower bounded by certain functions of the incompatibility of subsets of the available measurements. We prove the tightness of some of our bounds by providing explicit examples based on mutually unbiased bases. Finally, we discuss the consequences of our results for the nonlocality that can be gained by enlarging the number of measurements in a Bell experiment. 
\end{abstract}

\maketitle

\begin{figure}
\includegraphics[scale=0.55]{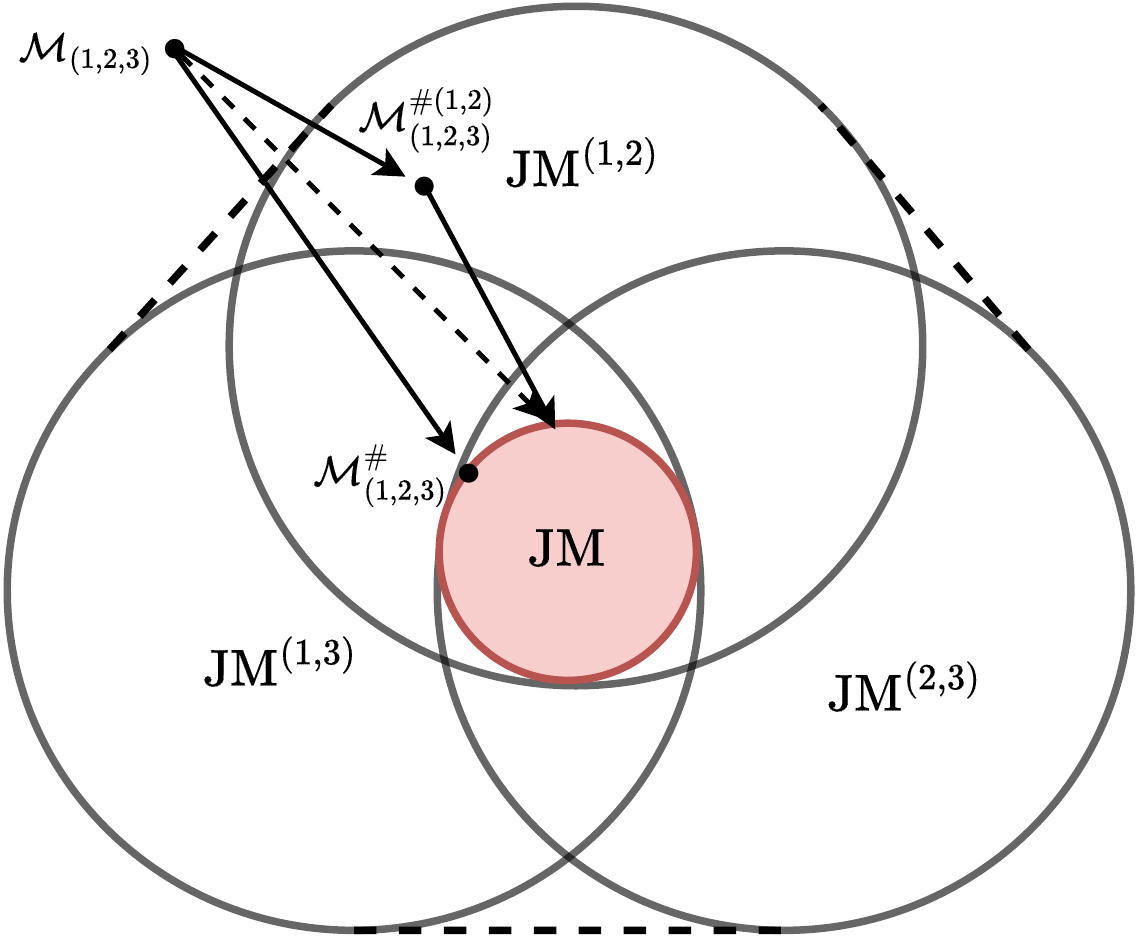}
 \caption{Different structures of incompatibility for three measurements, see also Ref.~\cite{PhysRevLett.123.180401}. The sets $\mathrm{JM}^{(s,t)}$ contain assemblages of measurements where the pairs $(s,t)$ are compatible. Their intersection $\mathrm{JM}^{\mathrm{pair}} \coloneqq \mathrm{JM}^{(1,2)} \cap \mathrm{JM}^{(1,3)} \cap \mathrm{JM}^{(2,3)}$ contains all pairwise compatible assemblages, with the set $\mathrm{JM}$ of all jointly measurable assemblages as a proper subset. Assemblages not contained in the convex hull $\mathrm{JM}^{\mathrm{conv}} \coloneqq \mathrm{Conv}(\mathrm{JM}^{(1,2)},\mathrm{JM}^{(1,3)}, \mathrm{JM}^{(2,3)})$ of the sets $\mathrm{JM}^{(s,t)}$, i.e., those that cannot be written as a convex combination of assemblages from the sets $\mathrm{JM}^{(1,2)}$, $\mathrm{JM}^{(1,3)}$, and $\mathrm{JM}^{(2,3)}$ are genuinely triplewise incompatible. The incompatibility of $\mathcal{M}_{(1,2,3)}$ is given by the distance to its closest jointly measurable approximation $\mathcal{M}^{\#}_{(1,2,3)}$. This distance can be upper bounded using the triangle inequality via the assemblage $\mathcal{M}^{\# (1,2)}_{(1,2,3)}$.}
  \label{IncompatiblityStructures}
\end{figure}

\begin{figure}
\includegraphics[scale=0.5]{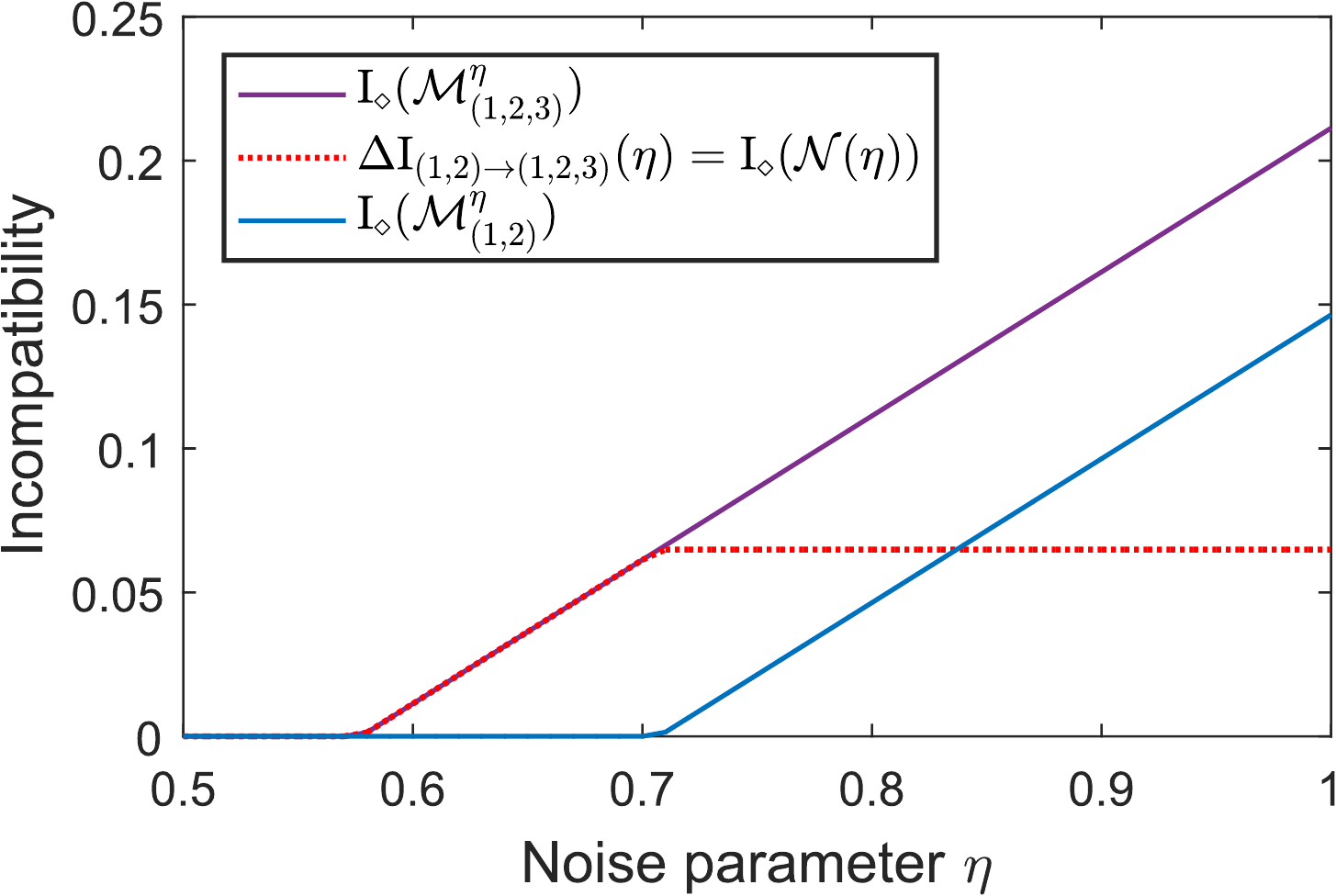}
\caption{Incompatibility gain for adding a third Pauli measurement. The gained incompatibility is given by the red (dotted) line. In the regime where $\Idiamond(\mathcal{M}_{(1,2)}) \neq 0$, the gained incompatibility remains constant. The red (dotted) curve and the blue curve add up to the violet one.}
\label{IncompatibilityGainFigure}
\end{figure}

The incompatibility of quantum measurements, i.e., the impossibility of measuring specific observable quantities simultaneously, is one of quantum physics' most prominent and striking properties. First discussed by Heisenberg \cite{Heisenberg1927} and Robertson~\cite{PhysRev.34.163}, this counterintuitive feature was initially thought of as a puzzling curiosity that represents a drawback for potential applications. Nowadays, measurement incompatibility~\cite{2112.06784,Heinosaari2016} is understood as a fundamental property of nature that lies at the heart of many quantum information processing tasks, such as quantum state discrimination~\cite{PhysRevLett.124.120401,PhysRevLett.122.130402,PhysRevLett.122.130404,Oszmaniec2019,PhysRevLett.125.110401,PhysRevLett.125.110402}, quantum cryptography~\cite{Bennett2014,Pirandola:20}, and quantum random access codes~\cite{Carmeli2020,PhysRevLett.125.080403}. Even more importantly, incompatible measurements are a crucial requirement for quantum phenomena such as quantum contextuality~\cite{2102.13036}, EPR-steering~\cite{RevModPhys.92.015001,Cavalcanti2016}, and Bell nonlocality~\cite{Nonlocality_review}. \\
\indent Its fundamental importance necessitates gaining a deep understanding of measurement incompatibility from a qualitative and quantitative perspective. By its very definition, measurement incompatibility arises when at least $m \geq 2$ measurements are considered that cannot be measured jointly by performing a single measurement instead. Generally, adding more measurements to a measurement scheme may allow for more incompatibility, hence increasing advantages in certain applications. \\
\indent However, it is unclear how much incompatibility can be gained from adding further measurements to an existing measurement scheme and on what this potential gain depends. Similarly, it is unclear how the incompatibility of measurement pairs contributes towards the total incompatibility of the whole set. Answering these questions is crucial to understanding specific protocols' power over others, such as protocols involving different numbers of \ac{MUB}~in quantum key distribution \cite{Bennett2014, PhysRevLett.81.3018}. While it is known~\cite{PhysRevLett.123.180401} that the different incompatibility structures (e.g., genuine triplewise and pairwise incompatibility) arising for $m \geq 3$ measurements set different limitations on the violation of Bell inequalities and incompatibility structures beyond two measurements have also been studied in~\cite{PhysRevA.89.052126,Heinosaari2008,Liang2011}, so far, no systematical way to quantify the gained advantage is known. \\
\indent The systematical and quantitative analysis of incompatibility structures in this work is inspired by the analysis of the distribution of multipartite entanglement \cite{PhysRevA.61.052306} and coherence \cite{PhysRevLett.116.150504}, leading to the observation that these quantum resources behave monogamously across subsets of systems. Despite the mathematical differences, our work follows physically a similar path by studying the distribution of quantum incompatibility across subsets of measurements. Namely, we show how an assemblage's incompatibility depends quantitatively on its subsets' incompatibilities. More specifically, we show how the potential gain of adding measurements to an existing measurement scheme is bounded by the incompatibility of the parent \acp{POVM} that approximate the respective subsets of measurements by a single measurement. \\
\indent Our results reveal the \emph{polygamous nature of measurement incompatibility} in the sense that an assemblage of more than two measurements can only be highly incompatible if all its subsets and the respective parent \acp{POVM} of the closest jointly measurable approximation of these subsets are highly non-jointly measurable. Our considerations lead to a new notion of measurement incompatibility that accounts only for a specific measurement's incompatibility contribution. We prove the relevance of our bounds on the incompatibility that can maximally be gained by showing that they are tight for particular measurement assemblages based on \ac{MUB}. Finally, we show that our results have direct consequences for steering and Bell nonlocality and discuss future applications of our results and methods. \\
\indent \textit{Preliminaries.}\textemdash We describe a quantum measurement most generally by a \ac{POVM}, i.e., a set $\lbrace M_a \rbrace$ of operators $0 \leq M_{a} \leq \mathds{1}$ such that $\sum_a M_a = \mathds{1}$. Given a state $\rho$, the probability of obtaining outcome $a$ is given by the Born rule $p(a) = \Tr[M_a \rho]$. A \emph{measurement assemblage} is a collection of different \acp{POVM} with operators $M_{a \vert x}$, where $x$ denotes the particular measurement. We write an assemblage $\mathcal{M}_{(1,2,\cdots,m)} = (\mathcal{M}_1, \mathcal{M}_2, \cdots, \mathcal{M}_m) $ of $m$ measurements as an ordered list of \acp{POVM}, where $\mathcal{M}_x$ refers to the $x$-th measurement. For instance, $\mathcal{M}_{(1,2,3)} = (\mathcal{M}_1, \mathcal{M}_2,\mathcal{M}_3)$ refers to an assemblage with three (different) measurements and $\mathcal{M}_{(1,2,2)} = (\mathcal{M}_1, \mathcal{M}_2,\mathcal{M}_2)$ denotes an assemblage where the second and the third \ac{POVM} are equal.  \\
\indent An assemblage $\mathcal{M}$ is called \emph{jointly measurable} if it can be simulated by a single \emph{parent \ac{POVM}} $\lbrace G_{\lambda} \rbrace$ and conditional probabilities $p(a \vert x, \lambda)$ such that 
\begin{align}
\label{IncompatiblityDef}
M_{a \vert x} = \sum_{\lambda} p(a \vert x, \lambda) G_{\lambda} \ \forall \ a,x,  
\end{align}
and it is called \emph{incompatible} otherwise. Here, we call $G(\mathcal{M})$ a parent \ac{POVM} of a jointly measurable assemblage $\mathcal{M}$. Various functions can quantify measurement incompatibility \cite{Pusey2015,Designolle2019,2207.05722}. The most suitable incompatibility quantifier for our purposes is the recently introduced \emph{diamond distance quantifier}~\cite{Tendick2023}, given by
\begin{align}
\label{Incompatibility}
\Idiamond(\mathcal{M}^\mathbf{p}) = \min\limits_{\mathcal{F} \in \mathrm{JM}} \sum_{x} p(x)  \Ddiamond(\Lambda_{\mathcal{M}_x}, \Lambda_{\mathcal{F}_x}),  
\end{align} 
where $\mathrm{JM}$ denotes the set of jointly measurable assemblages, $\Lambda_{\mathcal{M}_x} = \sum_a \mathrm{Tr}[M_{a \vert x} \rho] \vert a \rangle \langle a \vert$ is the \emph{measure-and-prepare channel} associated to the measurement $\mathcal{M}_x,$ and $\Ddiamond(\Lambda_1, \Lambda_2) = \max\limits_{\rho \in \mathcal{S}(\mathcal{H} \otimes \mathcal{H})} \dfrac{1}{2} \lVert ((\Lambda_1-\Lambda_2) \otimes \mathds{1}_{d})\rho \rVert_1$ is the \emph{diamond distance}~\cite{Kitaev2002} between two channels $\Lambda_1$ and $\Lambda_2$, with the trace norm $\lVert X \rVert_1 = \Tr[\sqrt{X^{\dagger}X}]$.  Furthermore, $\mathcal{M}^\mathbf{p} = (\mathcal{M}, \mathbf{p})$ denotes a \emph{weighted measurement assemblage}, where $\mathbf{p}$ contains the probabilities $p(x)$ with which measurement $x$ is performed. Note that $\Idiamond(\mathcal{M}^\mathbf{p})$ is induced by the general distance $\Ddiamond(\mathcal{M}^\mathbf{p}, \mathcal{N}^\mathbf{p}) \coloneqq \sum_x p(x) \Ddiamond(\Lambda_{\mathcal{M}_x}, \Lambda_{\mathcal{N}_x})$ between two assemblages $\mathcal{M}^\mathbf{p}$ and $\mathcal{N}^\mathbf{p}$. \\
\indent 
We denote by $\mathcal{M}^{\#}_{(1,2,\cdots,m)}$ the closest jointly measurable assemblage with respect to $\mathcal{M}_{(1,2,\dots,m)}$, i.e., the \emph{arg-min} on the RHS in Eq.~\eqref{Incompatibility}. While $\mathcal{M}^{\#}_{(1,2,\cdots,m)}$ and its underlying parent \ac{POVM} are generally not unique~\cite{Heinosaari2008,Guerini2018}, all the results derived in this work hold for any valid choice, as we do not assume uniqueness. If we only approximate a subset of $n < m$ measurements of $\mathcal{M}_{(1,2,\dots,m)}$ by jointly measurable ones, for instance the first $n$ settings, while keeping the remaining measurements unchanged, we write $\mathcal{M}^{\# (1,2,\dots,n)}_{(1,2,\cdots,m)}$. \\
\indent The diamond distance quantifier $\Idiamond(\mathcal{M}^\mathbf{p})$ \cite{Tendick2023} is particularly well-suited for our purposes, as it is not only monotonous under the application of quantum channels and classical simulations but it also inherits all properties of a distance (in particular the triangle inequality of $\Ddiamond(\mathcal{M}^\mathbf{p}, \mathcal{N}^\mathbf{p})$), and it is written in terms of a convex combination of the individual measurement's distances. \\
\indent Besides these technical requirements, the quantifier $\Idiamond(\mathcal{M}^\mathbf{p})$ admits the operational interpretation of average single-shot distinguishability of the assemblage $\mathcal{M}$ from its closest jointly measurable assemblage $\mathcal{M}^{\#}$. Furthermore, it can be used to upper bound the amount of steerability and nonlocality that can be revealed by the measurements $\mathcal{M}$ in Bell-type experiments \cite{Tendick2023}. \\
\indent For pedagogical reasons, we focus in the main text on the scenario $2 \rightarrow 3$, i.e., we consider an assemblage of $m=2$ measurements that is promoted to one with $m'=3$ settings. Furthermore, we set $p(x)$ to be uniformly distributed and simply use the symbol $\mathcal{M}$ for the weighted assemblage in this case. We refer to the \ac{SM}~\cite{Supplemental_Material} for all proofs, more background information, and generalizations to an arbitrary number of measurements and general probability distributions.  \\
\indent Adding a third measurement $\mathcal{M}_3$ to the assemblage $\mathcal{M}_{(1,2)} = (\mathcal{M}_1,\mathcal{M}_2)$ is mathematically described by the concatenation of ordered lists, using the symbol $\mdoubleplus$, i.e., we write 
\begin{align}
\label{Concatenation}
\mathcal{M}_{(1,2,3)} = \mathcal{M}_{(1,2)} \mdoubleplus \mathcal{M}_3 = (\mathcal{M}_1,\mathcal{M}_2,\mathcal{M}_3).
\end{align}
Using the concatenation of ordered lists, we formally define $\mathcal{M}^{\# (1,2)}_{(1,2,3)}$ such that
\begin{align}
\mathcal{M}^{\# (1,2)}_{(1,2,3)} \coloneqq \mathcal{M}^{\#}_{(1,2)} \mdoubleplus \mathcal{M}_3.
\end{align} 
\indent Three measurements allow for incompatibility structures \cite{PhysRevA.89.052126,Heinosaari2008,Liang2011,PhysRevLett.123.180401} beyond Eq.~\eqref{IncompatiblityDef}. We define the sets $\mathrm{JM}^{(s,t)}$ with $s \neq t \in \lbrace 1,2,3 \rbrace$ as those containing assemblages in which the measurements $s$ and $t$ are jointly measurable.
This allows us to define \emph{pairwise} and \emph{genuinely triplewise incompatible} assemblages \cite{PhysRevLett.123.180401} as those that are \emph{not} contained in the intersection and the convex hull of the sets $\mathrm{JM}^{(s,t)}$, respectively. See also Figure~\ref{IncompatiblityStructures} for a graphical representation and more details. 
\\
\indent \textit{Incompatibility gain.}\textemdash We investigate the \emph{incompatibility gain} obtained from adding measurements to an already available assemblage. That is, for an assemblage $\mathcal{M}_{(1,2,3)}$ defined via Eq.~\eqref{Concatenation} we want to quantify the gain
\begin{align}
\label{IncompatibilityGain}
\Delta\mathrm{I}_{(1,2) \rightarrow (1,2,3)} \coloneqq \Idiamond(\mathcal{M}_{(1,2,3)})-\Idiamond(\mathcal{M}_{(1,2)}).
\end{align}
\indent Note that $\Delta\mathrm{I}_{(1,2) \rightarrow (1,2,3)}$ is the difference of two quantities that can be computed via \acp{SDP}~\cite{Tendick2023}, however, the purely numerical value of the gained incompatibility does only provide limited physical insights by itself. While it seems generally challenging to find an exact analytical expression for the incompatibility gain, we will derive bounds on it in the following.\\
\indent Our approach relies on a two-step protocol. First, we employ a \emph{measurement splitting}, i.e., instead of considering the incompatibility of $\mathcal{M}_{(1,2,3)}$, we consider the incompatibility  $\Idiamond(\mathcal{M}_{(1,2,1,3,2,3)})$. That is, each measurement of $\mathcal{M}_{(1,2,3)}$ is now split up into two equivalent ones, each occurring with a probability of $\tfrac{1}{6}$. 
% (we show in Section II of the \ac{SM} \cite{Supplemental_Material} that biased splittings i.e., unequal probabilities summing to $\tfrac{1}{3}$ can also be used)
Furthermore, it holds $\Idiamond(\mathcal{M}_{(1,2,3)}) = \Idiamond(\mathcal{M}_{(1,2,1,3,2,3)})$ since the assemblages can be converted into each other by (reversible) classical post-processing~\cite{Supplemental_Material} (Section II). The second step involves a particular instance of the triangle inequality and uses specifically that $\Idiamond(\mathcal{M})$ is defined as convex combination over the individual settings. More precisely, let 
\begin{align}
\label{Definition_N}
\mathcal{N}= \mathcal{M}^{\#}_{(1,2)} \mdoubleplus \mathcal{M}^{\#}_{(1,3)} \mdoubleplus \mathcal{M}^{\#}_{(2,3)},
\end{align} be an assemblage that contains itself three assemblages (of two measurements each) that are the closest jointly measurable approximations with respect to the individual subsets of $\mathcal{M}_{(1,2,3)}$. We point out that $\mathcal{N}$ itself can be incompatible in general. Using the triangle inequality, it follows that
\begin{align}
\Idiamond(\mathcal{M}_{(1,2,3)}) &= \Idiamond(\mathcal{M}_{(1,2,1,3,2,3)}) \\
&\leq  \Ddiamond(\mathcal{M}_{(1,2,1,3,2,3)}, \mathcal{N}) +\Idiamond(\mathcal{N}). \nonumber
\end{align}
Due to our choice of $\mathcal{N}$, the term $\Ddiamond(\mathcal{M}_{(1,2,1,3,2,3)}, \mathcal{N})$ evaluates to the average incompatibility of the subsets, as we can split the sum over all six settings into three pairs, i.e. we obtain
\begin{align}
\label{BoundIncompatibilityAverage}
\Idiamond(\mathcal{M}_{(1,2,3)}) &\leq \dfrac{1}{3}\big[\Idiamond(\mathcal{M}_{(1,2)})+\Idiamond(\mathcal{M}_{(1,3)}) \\
&+\Idiamond(\mathcal{M}_{(2,3)})\big] + \Idiamond(\mathcal{N}). \nonumber
\end{align}
That is, the incompatibility of $\mathcal{M}_{(1,2,3)}$ is upper bounded by the average incompatibility of its two-measurement subsets plus the incompatibility  $ \Idiamond(\mathcal{N})$ that contains the information about how incompatible the respective closest jointly measurable \acp{POVM} are with each other. Notice that  $\Idiamond(\mathcal{N}) \leq \Idiamond(\mathcal{G})$ holds, where 
\begin{align}
\label{Definition_G}
\mathcal{G} = G(\mathcal{M}^{\#}_{(1,2)}) \mdoubleplus G(\mathcal{M}^{\#}_{(1,3)}) \mdoubleplus G(\mathcal{M}^{\#}_{(2,3)}) 
\end{align} is the assemblage that contains the parent \acp{POVM} $G$ of the respective subsets, as $\mathcal{N}$ is a classical post-processing of $\mathcal{G}$~\cite{Supplemental_Material} (Section II). This shows that the incompatibility of $\mathcal{M}_{(1,2,3)}$ is limited on two different levels through its subsets. Moreover, it reveals a type of \emph{polygamous} behavior of incompatibility. For high incompatibility of $\mathcal{M}_{(1,2,3)}$ each of the subsets, as well as the underlying parent \acp{POVM} of the respective jointly measurable approximations, have to be highly incompatible. Coming back to the incompatibility gain, we are ready to present our first main result. 
\begin{result}
\label{Result1}
Let $\Idiamond(\mathcal{M}_{(1,2)})~\geq~\max \lbrace \Idiamond(\mathcal{M}_{(1,3)}), \Idiamond(\mathcal{M}_{(2,3)}) \rbrace$. It follows that the incompatibility gain as defined in Eq.~\eqref{IncompatibilityGain} is bounded such that
\begin{align}
\label{BoundIncomatibilityGain}
\Delta\mathrm{I}_{(1,2) \rightarrow (1,2,3)} \leq  \Idiamond(\mathcal{N}) \leq \Idiamond(\mathcal{G}). \end{align}
\end{result}
This means that the potential incompatibility gain is limited by the incompatibility of the assemblage $\mathcal{N}$ in Eq.~\eqref{Definition_N}, i.e., the concatenation of the respective closest jointly measurable approximations of the subsets. Physically more intuitive, it is limited by the incompatibility of the assemblage that contains the respective parent \acp{POVM}. The assumption $\Idiamond(\mathcal{M}_{(1,2)}) \geq \max \lbrace \Idiamond(\mathcal{M}_{(1,3)}), \Idiamond(\mathcal{M}_{(2,3)}) \rbrace$ represents no loss of generality for all practical purposes, as one can simply optimize over all possible two-measurement subsets. \\
\indent We show in the \ac{SM} \cite{Supplemental_Material} that Result \ref{Result1} can be generalized to 
\begin{align}
\Delta\mathrm{I}_{(1,\cdots,m) \rightarrow (1,\cdots,m,m+1)} \leq  \Idiamond(\mathcal{N}) \leq \Idiamond(\mathcal{G}),   
\end{align}
by appropriately redefining $\mathcal{N}$ and $\mathcal{G}$. \\
\indent We point out that Result~\ref{Result1} allows for the definition of a single \emph{maximally incompatible additional} measurement, in the sense that it is the measurement $\mathcal{M}_3$ that maximizes the incompatibility gain $\Delta\mathrm{I}_{(1,2) \rightarrow (1,2,3)}$ for a given assemblage $\mathcal{M}_{(1,2)}$. As an illustrative example, we consider the three projective measurements $\lbrace \Pi_{a \vert x} \rbrace$ which represent the Pauli $X,Y,Z$ observables subjected to white noise, i.e., we analyze the incompatibility of the assemblage $\mathcal{M}^{\eta}_{(1,2,3)}=(\mathcal{M}^{\eta}_1,\mathcal{M}^{\eta}_2,\mathcal{M}^{\eta}_3)$ defined via
\begin{align}
\label{DepolarizedPaulis}
M^{\eta}_{a \vert x} = \eta \Pi_{a \vert x} + (1-\eta) \Tr[\Pi_{a \vert x}] \dfrac{\mathds{1}}{2}, 
\end{align}
where $(1-\eta)$ is the noise level. It holds in this particular case that (see Figure~\ref{IncompatibilityGainFigure}):  
\begin{align}
\Delta\mathrm{I}_{(1,2) \rightarrow (1,2,3)}(\eta) =  \Idiamond(\mathcal{N}(\eta)),    
\end{align}
which we prove analytically in the \ac{SM}~\cite{Supplemental_Material} (Section VI). For the regime $\tfrac{1}{\sqrt{2}} \leq \eta \leq 1$ we 
also show that $\Idiamond(\mathcal{N} (\eta)) = \Idiamond(\mathcal{M}^{1/\sqrt{2}}_{(1,2,3)})$, which means that the gained incompatibility is exactly given by the incompatibility of $\mathcal{M}^{\eta}_{(1,2,3)}$ at the noise threshold where it becomes \emph{pairwise compatible}. \\
\indent Our methods can also be applied to obtain lower bounds. For instance, we show \cite{Supplemental_Material} (Section III) that $\Idiamond(\mathcal{M}_{(1,2,3)})$ is bounded by the average subset incompatibility: 
\begin{align}
\label{LowerBoundAverage}
\Idiamond(\mathcal{M}_{(1,2,3)}) \geq \dfrac{1}{3}[\Idiamond(\mathcal{M}_{(1,2)})+\Idiamond(\mathcal{M}_{(1,3)}) +\Idiamond(\mathcal{M}_{(2,3)})].
\end{align}
In general, $\Idiamond(\mathcal{M}_{(1,2,3)}) < \Idiamond(\mathcal{M}_{(1,2)}$ is possible, i.e., adding a measurement to an assemblage can actually decrease the incompatibility, if we do \emph{not} optimize over the input distribution $\mathbf{p}$. For instance, adding a measurement $\mathcal{M}_{3} $ that is jointly measurable with $ \mathcal{M}^{\#}_{(1,2)}$, such as an identity measurement, generally decreases the incompatibility.\\
\indent Another way to see how the incompatibility of an assemblage $\mathcal{M}_{(1,2,3)}$ can be upper bounded in terms of the incompatibility $\Idiamond(\mathcal{M}_{(1,2)})$ plus the gained incompatibility due to measurement $\mathcal{M}_3$ relies on directly applying specific instances of the triangle inequality without splitting the measurements. \\
\indent \textit{A new notion of incompatibility.}\textemdash Consider the general assemblage $\mathcal{M}_{(1,2,3)}$ as defined in Eq.~\eqref{Concatenation}. Due to the triangle inequality, see also Figure~\ref{IncompatiblityStructures}, it holds
\begin{align}
\Idiamond(\mathcal{M}_{(1,2,3)}) \leq \Ddiamond(\mathcal{M}_{(1,2,3)}, \mathcal{N}_{(1,2,3)}) + \Idiamond(\mathcal{N}_{(1,2,3)}),
\end{align}
for any assemblage $\mathcal{N}_{(1,2,3)}$. By choosing $\mathcal{N}_{(1,2,3)} = \mathcal{M}^{\# (1,2)}_{(1,2,3)} \coloneqq \mathcal{M}^{\#}_{(1,2)} \mdoubleplus \mathcal{M}_3$, we obtain our second main result. \begin{result}
Let $\mathcal{M}_{(1,2,3)} = \mathcal{M}_{(1,2)} \mdoubleplus \mathcal{M}_3$ be a concatenated measurement assemblage and $\mathcal{M}^{\#}_{(1,2)}$ the closest jointly measurable approximation of $\mathcal{M}_{(1,2)}$. It holds 
\begin{align}
\label{IncompatibilitySimpleUpperBound}
\Idiamond(\mathcal{M}_{(1,2,3)}) \leq \dfrac{2}{3} \Idiamond(\mathcal{M}_{(1,2)}) + 
\Idiamond(\mathcal{M}^{\# (1,2)}_{(1,2,3)}).
\end{align}
\end{result}
This means that the incompatibility of $\mathcal{M}_{(1,2,3)}$ is upper bounded by the incompatibility of the subset $\mathcal{M}_{(1,2)}$, weighted with the probability $p = \tfrac{2}{3}$, plus the incompatibility of the added measurement $\mathcal{M}_3$ with the closest jointly measurable approximation $\mathcal{M}^{\#}_{(1,2)}$ of $\mathcal{M}_{(1,2)}$. In \cite{Supplemental_Material} (Section III) we also show that the incompatibility of $\mathcal{M}_{(1,2,3)}$ is lower bounded by
\begin{align}
\label{LowerBoundSimple}
\Idiamond(\mathcal{M}_{(1,2,3)}) \geq \tfrac{2}{3}\Idiamond(\mathcal{M}_{(1,2)}).    
\end{align}
\indent The only incompatibility that contributes to $\Idiamond(\mathcal{M}^{\# (1,2)}_{(1,2,3)})$ is the incompatibility of $\mathcal{M}_3$ with the assemblage $\mathcal{M}^{\#}_{(1,2)}$, which itself is jointly measurable. Therefore, this term in Eq. \eqref{IncompatibilitySimpleUpperBound} can be understood as a new notion of incompatibility of the assemblage $\mathcal{M}_{(1,2,3)}$, where all incompatibilities apart of the contribution that comes from the presence of measurement $\mathcal{M}_3$ are omitted. \\
\indent  We show analytically in the \ac{SM}~\cite{Supplemental_Material} (Section VI) that the bound in Eq.~\eqref{IncompatibilitySimpleUpperBound} is tight for depolarized Pauli measurements (see Eq.~\eqref{DepolarizedPaulis}). Moreover, we show analytically that a similar bound is tight for certain measurements based on $d$-dimensional \ac{MUB} in cases where the number of measurements $m$ is changed such that $m=2\rightarrow m'=d$, $m=2\rightarrow m'=d+1$, and $m=d\rightarrow m'=d+1$. Namely, we prove and analyze the generalization of Eq.~\eqref{IncompatibilitySimpleUpperBound}:
\begin{align}
\Idiamond(\mathcal{M}_{(1,2,\cdots,m)}) \leq \dfrac{\lvert C \vert}{m} \Idiamond(\mathcal{M}_C) + \Idiamond(\mathcal{M}^{\# C}_{(1,2,\cdots,m)}),
\end{align}
for any assemblage $\mathcal{M}_{(1,2,\cdots,m)}$ and any subset $C$ of measurements with cardinality $\lvert C \vert$.\\
\indent \textit{Incompatibility decomposition.}\textemdash Looking at the results in Figure~\ref{IncompatibilityGainFigure} leads to the question of whether there exists a more general decomposition of $\Idiamond(\mathcal{M}_{(1,2,3)})$ into different incompatibility structures. Indeed, since $\Idiamond(\mathcal{M})$ is a \emph{distance-based} incompatibility quantifier, our final main result follows.
\begin{result}
\label{Result3}
The incompatibility of any assemblage $\mathcal{M}$ of $m=3$ measurements is upper bounded such that
\begin{align}
\label{IncompatiblityDecomposition}
\Idiamond(\mathcal{M}) &\leq \Idiamond^{\mathrm{gen}}(\mathcal{M}) + \Idiamond^{\mathrm{pair}}(\mathcal{M})+\Idiamond^{\mathrm{hol}}(\mathcal{M}),
\end{align}
where $\Idiamond^{\mathrm{gen}}(\mathcal{M})$ is the \emph{genuine triplewise incompatibility} of $\mathcal{M}$, i.e., its minimal distance to an assemblage $\mathcal{M}^{\mathrm{conv}} \in \mathrm{JM}^{\mathrm{conv}}$. Furthermore, we define $\Idiamond^{\mathrm{pair}}(\mathcal{M}) \coloneqq \Ddiamond(\mathcal{M}^{\mathrm{conv}},\mathcal{M}^{\mathrm{pair}})$ 
to be the \emph{pairwise incompatibility}, 
where $\mathcal{M}^{\mathrm{pair}} \in \mathrm{JM}^{\mathrm{pair}}$ is the closest pairwise compatible assemblage with respect to $\mathcal{M}^{\mathrm{conv}}$. 
We call the term $\Idiamond^{\mathrm{hol}}(\mathcal{M}) \coloneqq \Idiamond(\mathcal{M}^{\mathrm{pair}})$ the 
\emph{hollow incompatibility}, which implicitly depends on $\mathcal{M}$, see also Figure~\ref{IncompatiblityStructures} and Ref.~\cite{PhysRevLett.123.180401}. 
\end{result}
We emphasize that the bound in Eq.~\eqref{IncompatiblityDecomposition} relies crucially on the distance properties of the quantifier $\Idiamond(\mathcal{M})$ and \emph{cannot} be adapted directly to robustness or weight quantifiers~\cite{Pusey2015,Designolle2019}. In the \ac{SM}~\cite{Supplemental_Material} (Section VII) we show that the decomposition in Eq.~\eqref{IncompatiblityDecomposition} is tight for the three Pauli measurements, and give numerical indication that this is generally the case for measurements based on \ac{MUB}. \\
\indent \textit{Implications for steering and Bell nonlocality.}\textemdash Due to the mathematical structure of our methods, they can directly be applied to quantum steering and Bell nonlocality. Note that both of these phenomena occur in a scenario that is similar to the one for measurement incompatibility. Namely, they depend on the properties of a set of at least two measurements, while a single measurement by itself does not contain any resource. This distinguishes the above concepts from resource theories of single \acp{POVM} (see, e.g. \cite{PhysRevLett.119.190501,PhysRevLett.122.140403,Oszmaniec2019}) where the resource gain can trivially be determined by considering averages of single POVM resources \cite{Tendick2023}. We describe our results regarding steering and nonlocality in more detail in the \ac{SM}~\cite{Supplemental_Material} (Section IV). The analysis of the gain in nonlocal correlations in Bell experiments is particularly interesting as it seems fundamentally different from incompatibility and steering. Consider a Bell experiment where Alice performs $m_A=3$ and Bob $m_B=2$ measurements.
%Analogous bounds to Eq.~\eqref{BoundIncompatibilityAverage} (\ac{SM} \cite{Supplemental_Material}, Section IV) for the nonlocality of the distribution $\mathbf{q}_{(1,2,3)}$ (where the indicies refers to Alice's settings) involve the nonlocality that can be obtained from all the two-measurement subsets of Alice.
Focusing on dichotomic measurements, we observe the following intriguing effect: Alice cannot find three measurements, such that the three \ac{CHSH} inequalities \cite{PhysRevLett.23.880} $\mathrm{CHSH}_{(i,j)} \coloneqq \langle A_i \otimes B_1 \rangle + \langle A_i \otimes B_2 \rangle + \langle A_j \otimes B_1 \rangle - \langle A_j \otimes B_2 \rangle \leq 2$ with $(i,j) \in \lbrace (1,2), (1,3), (2,3) \rbrace$ are simultaneously maximally violated. That means,  $\mathrm{CHSH}_{(1,2,3)} \coloneqq \tfrac{1}{3}( \mathrm{CHSH}_{(1,2)}+\mathrm{CHSH}_{(1,3)}+\mathrm{CHSH}_{(2,3)}) \leq  \tfrac{4\sqrt{2}+2}{3} < 2\sqrt{2}$ holds in quantum theory. This implies, that the average two-subset nonlocality is lower than the maximal obtainable nonlocality with two measurements on Alice's side. 
%This effect seems to be particular to nonlocality and distinguishes it from steering and incompatibility. 
%Therefore, this observation could prove crucial in understanding why nonlocality seems to behave differently to incompatibility and steering, in the sense that using more than two measurements is not known to provide any advantages for the maximal obtainable Bell nonlocality~\cite{Arajo2020,PhysRevA.97.022111}. }\\
\\
\indent \textit{Conclusion and outlook.}\textemdash In this work, we analyzed how much incompatibility can maximally be gained by adding measurements to an existing measurement scheme. We showed that this gain is upper bounded by the incompatibility of the underlying parent \acp{POVM} that approximate subsets of measurements. We proved the relevance of our bounds analytically by showing that they are tight for specific measurements based on \ac{MUB}. Moreover, we showed that our methods are directly applicable to quantum steering and Bell nonlocality. For nonlocality specifically, we discovered a promising path to understand better why using more than two measurements may not provide any advantage for maximal nonlocal correlations \cite{Arajo2020,PhysRevA.97.022111}. Our results reveal the polygamous nature of distributed quantum incompatibility, in stark contrast to the monogamy of entanglement \cite{PhysRevA.61.052306} and coherence \cite{PhysRevLett.116.150504} across subsystems of multipartite quantum states. While we focused in this text on $m=3$ measurements, all our findings, in particular, Results \ref{Result1}-\ref{Result3} can be generalized to an arbitrary number of measurements $m$ (see \cite{Supplemental_Material}, Section V).  \\
\indent Our work provides a foundation for several new directions of research. While we focused on a particular distance-based quantifier here, the alternative distance-based quantifiers proposed in \cite{Tendick2023} do also possess the necessary properties to be used in a similar way. It would be interesting to see whether resource quantifiers such as the incompatibility robustness~\cite{Designolle2019} or weight~\cite{Pusey2015} can also be used to analyze how the incompatibility of an assemblage depends on its subsets. Our methods might also prove helpful to find better bounds on the incompatibility of general assemblages and particularly maximally assemblages. Finally, it would be interesting to analyze the performance gain of specific cryptography \cite{PhysRevLett.81.3018,Bennett2014} or estimation protocols \cite{2206.08912} with different numbers of measurements. \\

\begin{acknowledgments}
We thank Thomas Cope, Federico Grasselli, Martin Kliesch, Nikolai Miklin, Martin Pl\'{a}vala, Isadora Veeren, Thomas Wagner, and Zhen-Peng Xu for helpful discussions.
This research was partially supported
by the EU H2020 QuantERA ERA-NET Cofund in
Quantum Technologies project QuICHE, and by the Federal Ministry of Education and Research (BMBF)  within the funding program \grqq{}Forschung Agil - Innovative Verfahren f\"{u}r Quantenkommunikationsnetze\grqq{} in the joint project QuKuK (grant
number 16KIS1619).
\\[1em]
\end{acknowledgments}

\begin{acronym}[CGLMP]\itemsep 1\baselineskip
\acro{AGF}{average gate fidelity}
\acro{AMA}{associated measurement assemblage}

\acro{BOG}{binned outcome generation}

\acro{CGLMP}{Collins-Gisin-Linden-Massar-Popescu}
\acro{CHSH}{Clauser-Horne-Shimony-Holt}
\acro{CP}{completely positive}
\acro{CPT}{completely positive and trace preserving}
\acro{CPTP}{completely positive and trace preserving}
\acro{CS}{compressed sensing} 

\acro{DFE}{direct fidelity estimation} 
\acro{DM}{dark matter}

\acro{GST}{gate set tomography}
\acro{GUE}{Gaussian unitary ensemble}

\acro{HOG}{heavy outcome generation}

\acro{JM}{jointly measurable}

\acro{LHS}{local hidden-state model}
\acro{LHV}{local hidden-variable model}
\acro{LOCC}{local operations and classical communication}

\acro{MBL}{many-body localization}
\acro{ML}{machine learning}
\acro{MLE}{maximum likelihood estimation}
\acro{MPO}{matrix product operator}
\acro{MPS}{matrix product state}
\acro{MUB}{mutually unbiased bases} 
\acro{MW}{micro wave}

\acro{NISQ}{noisy and intermediate scale quantum}

\acro{POVM}{positive operator valued measure}
\acro{PVM}{projector-valued measure}

\acro{QAOA}{quantum approximate optimization algorithm}
\acro{QML}{quantum machine learning}
\acro{QMT}{measurement tomography}
\acro{QPT}{quantum process tomography}
\acro{QRT}{quantum resource theory}
\acroplural{QRT}[QRTs]{Quantum resource theories}

\acro{RDM}{reduced density matrix}

\acro{SDP}{semidefinite program}
\acro{SFE}{shadow fidelity estimation}
\acro{SIC}{symmetric, informationally complete}
\acro{SM}{Supplemental Material}
\acro{SPAM}{state preparation and measurement}

\acro{RB}{randomized benchmarking}
\acro{rf}{radio frequency}

\acro{TT}{tensor train}
\acro{TV}{total variation}

\acro{UI}{uninformative}

\acro{VQA}{variational quantum algorithm}

\acro{VQE}{variational quantum eigensolver}

\acro{WMA}{weighted measurement assemblage}

\acro{XEB}{cross-entropy benchmarking}

\end{acronym}

\bibliography{bibliography_2.bib}

\newpage

\onecolumngrid

\section*{Supplemental Material for "Distributed quantum incompatibility"}

In this Supplemental Material, we give detailed background information on measurement incompatibility, provide proofs for the results and statements in the main text, and discuss how to apply our results to steering and nonlocality. Furthermore, we show how to generalize our results to general sets of $m$ measurements and weighted measurement assemblages $\mathcal{M}^{\mathbf{p}} = (\mathcal{M},\mathbf{p})$ with general probability distributions $\mathbf{p}$.   

\section{Background information on incompatibility}

Here, we give detailed background information on the important properties of the diamond distance quantifier $\Idiamond(\mathcal{M}^\mathbf{p})$ defined in Eq.~(\textcolor{blue}{$2$}) in the main text. To provide a relatively self contained overview in this Supplemental Material, we also repeat the relevant definitions from the main text. An assemblage $\mathcal{M}_{(1,2,\cdots,m)} = (\mathcal{M}_1, \mathcal{M}_2, \cdots, \mathcal{M}_m)$ of $m$ measurements with outcomes $a$ and settings $x$ is called \emph{jointly measurable} if it can be simulated by a single \emph{parent \ac{POVM}} $\lbrace G_{\lambda} \rbrace$ and conditional probabilities $p(a \vert x, \lambda)$ such that 
\begin{align}
\label{Parent_POVM}
M_{a \vert x} = \sum_{\lambda} p(a \vert x, \lambda) G_{\lambda} \ \forall \ a,x,    
\end{align}
and it is called \emph{incompatible} otherwise. Note that the probabilities $p(a \vert x, \lambda)$ can always be identified with deterministic response functions $v(a \vert x, \lambda)$ since the randomness in $p(a \vert x, \lambda)$ can be shifted to the parent \ac{POVM} by appropriately redefining the $G_{\lambda}$. Let us denote by $\mathrm{JM}$ the set of all jointly measurable assemblages. For more than two measurements, there exist different sub-structures of incompatibility. Focusing on the case of three measurements, we define the sets $\mathrm{JM}^{(s,t)}$ with $s,t \in \lbrace 1,2,3 \rbrace$ such that $s \neq t$ as the sets containing assemblages in which the measurement $s$ and $t$ are jointly measurable. Their intersection $\mathrm{JM}^{\mathrm{pair}} \coloneqq \mathrm{JM}^{(1,2)} \cap \mathrm{JM}^{(1,3)} \cap \mathrm{JM}^{(2,3)}$ contains all assemblages in which any pair of two measurements are compatible, the so-called \emph{pairwise} compatible assemblages. On the other hand, the set $\mathrm{JM}^{\mathrm{conv}} \coloneqq \mathrm{Conv}(\mathrm{JM}^{(1,2)},\mathrm{JM}^{(1,3)}, \mathrm{JM}^{(2,3)})$ describes the convex hull of the sets $\mathrm{JM}^{(1,2)}$, $\mathrm{JM}^{(1,3)}$, and $\mathrm{JM}^{(2,3)}$, i.e., it contains all assemblage that can be written as a convex combination of assemblages where one pair of measurements is compatible. More formally, it contains all assemblages of the form
\begin{align}
\mathcal{M}_{(1,2,3)} = p_{(1,2)} \mathcal{J}^{(1,2)}_{(1,2,3)} + p_{(1,3)} \mathcal{J}^{(1,3)}_{(1,2,3)}+p_{(2,3)} \mathcal{J}^{(2,3)}_{(1,2,3)},
\end{align}
where $\mathcal{J}^{(s,t)}_{(1,2,3)} \in \mathrm{JM}^{(s,t)}$ and the convex combination is to be understood on the level of the individual \ac{POVM} effects. Finally an assemblage $\mathcal{M}_{(1,2,3)} \notin \mathrm{JM}^{\mathrm{conv}}$ is said to be \emph{genuinely triplewise incompatible}. Note, these notions can straightforwardly be generalized to more than three measurements. See also~\cite{PhysRevLett.123.180401} and for a graphical representation Figure~\textcolor{blue}{1} in the main text. \\
\indent To quantify the incompatibility as a resource, we use the \emph{diamond distance quantifier}~\cite{Tendick2023} given by
\begin{align}
\label{IncompatibilitySuppl}
\Idiamond(\mathcal{M}^\mathbf{p}) = \min\limits_{\mathcal{F} \in \mathrm{JM}} \sum_{x} p(x)  \Ddiamond(\Lambda_{\mathcal{M}_x}, \Lambda_{\mathcal{F}_x}), \end{align} 
where $\Lambda_{\mathcal{M}_x} = \sum_a \mathrm{Tr}[M_{a \vert x} \rho] \vert a \rangle \langle a \vert$ is the \emph{measure-and-prepare channel} associated to the measurement $\mathcal{M}_x,$ and $\Ddiamond(\Lambda_1, \Lambda_2) = \max\limits_{\rho \in \mathcal{S}(\mathcal{H} \otimes \mathcal{H})} \dfrac{1}{2} \lVert ((\Lambda_1-\Lambda_2) \otimes \mathds{1}_{d})\rho \rVert_1$ is the \emph{diamond distance}~\cite{Kitaev2002} between two channels $\Lambda_1$, and $\Lambda_2$, with the trace norm$\lVert X \rVert_1 = \Tr[\sqrt{X^{\dagger}X}]$. Technically speaking, $\Idiamond(\mathcal{M}^\mathbf{p})$ quantifies the incompatibility of a \emph{weighted assemblage} $\mathcal{M}^\mathbf{p} = (\mathcal{M}, \mathbf{p})$ which contains the information about the probabilities $p(x)$ with which the measurement $x$ is performed. The distance between two assemblages $\mathcal{M}^\mathbf{p}$ and $\mathcal{N}^\mathbf{p}$ that induces the quantifier $\Idiamond(\mathcal{M}^\mathbf{p})$ is given by 
\begin{align}
 \Ddiamond(\mathcal{M}^\mathbf{p}, \mathcal{N}^\mathbf{p}) \coloneqq \sum_x p(x) \Ddiamond(\Lambda_{\mathcal{M}_x}, \Lambda_{\mathcal{N}_x}).   
\end{align}
Like in the main text, we will simply write $\Idiamond(\mathcal{M})$ to imply the case where $p(x) = \tfrac{1}{m} \forall x$. We denote by $\mathcal{M}^{\#}_{(1,2,\cdots,m)}$ the closest jointly measurable assemblage to $\mathcal{M}_{(1,2,\dots,m)}$, i.e., the \emph{arg-min} on the RHS in Eq.~\eqref{IncompatibilitySuppl}. Therefore, $\mathcal{M}^{\#}_{(1,2,\cdots,m)}$ can be seen as the closest jointly measurable approximation of the assemblage $\mathcal{M}_{(1,2,\dots,m)}$. If we only approximate a subset of $n < m$ measurements of $\mathcal{M}_{(1,2,\dots,m)}$ by jointly measurable measurements, for instance the first $n$ settings, while keeping the remaining measurements unchanged, we write $\mathcal{M}^{\# (1,2,\dots,n)}_{(1,2,\cdots,m)}$. Adding measurements $\mathcal{M}'_{(m+1,m+2,\dots,m+n)} = (\mathcal{M}'_{m+1},\mathcal{M}'_{m+2},\cdots,\mathcal{M}'_{m+n})$ to the assemblage $\mathcal{M}_{(1,2,\cdots,m)} = (\mathcal{M}_1,\mathcal{M}_2,\cdots,\mathcal{M}_m)$ is mathematically described by the concatenation of ordered list, using the symbol $\mdoubleplus$, i.e., we write 
\begin{align}
\label{ConcatenationSuppl}
\mathcal{M}_{(1,2,\cdots,n+m)} = \mathcal{M}_{(1,2,\cdots,m)} \mdoubleplus \mathcal{M}'_{(m+1,m+2,\cdots,m+n)} = (\mathcal{M}_1,\mathcal{M}_2,\cdots,\mathcal{M}_m,\mathcal{M}'_{m+1},\cdots,\mathcal{M}'_{m+n}).
\end{align}
Using the notion of concatenation of ordered lists, we formally define
\begin{align}
\mathcal{M}^{\# (1,2,\dots,n)}_{(1,2,\cdots,m)} \coloneqq \mathcal{M}^{\#}_{(1,2,\cdots,n)} \mdoubleplus \mathcal{M}_{n+1} \mdoubleplus \cdots \mdoubleplus \mathcal{M}_{m}.    
\end{align}
\indent The diamond distance quantifier in Eq.~\eqref{IncompatibilitySuppl} is a \emph{faithful} resource quantifier, i.e., it holds that
\begin{align}
\Idiamond(\mathcal{M}^\mathbf{p}) = 0 \iff  \mathcal{M} = \mathcal{M}^{\#} \in \mathrm{JM}.
\end{align}
For the above statement to be true, we assume that $p(x) \neq 0 \ \forall x$, which is no  restriction, since measurements that are never performed can be excluded from the assemblage before calculating the incompatibility. \\
 \indent Furthermore, $\Idiamond(\mathcal{M}^\mathbf{p})$ is a monotone under any unital quantum channel $\Lambda^{\dagger}$ (these are exactly those channels that map \acp{POVM} to \acp{POVM}), i.e.,
\begin{align}
\Idiamond(\mathcal{M}^\mathbf{p}) \geq  \Idiamond(\Lambda^{\dagger}(\mathcal{M})^\mathbf{p}),
\end{align}
which follows from the fact that the trace distance is contractive under the application of \ac{CPTP} maps. 
Note that in the resource theory of incompatibility, all unital quantum channels $\Lambda^{\dagger}$ are free. Indeed, it is straight forward to see that $\lbrace \Lambda^{\dagger}(G_{\lambda}) \rbrace$ is a parent \ac{POVM} for the assemblage $\Lambda^{\dagger}(\mathcal{M})$ whenever $\lbrace G_{\lambda} \rbrace$ is a parent \ac{POVM} for $\mathcal{M}$. That is, it holds
\begin{align}
\Lambda^{\dagger}(M_{a \vert x}) = \sum_{\lambda} p(a \vert x, \lambda) \Lambda^{\dagger}(G_{\lambda}).    
\end{align}
\indent Additionally, $\Idiamond(\mathcal{M}^\mathbf{p})$ is non-increasing under classical simulations $\mathcal{M}' = \xi(\mathcal{M})$ with 
\begin{align}
\label{MeasurementSimulation}
M'_{b \vert y} = \sum_x p(x \vert y) \sum_a q(b \vert y,x,a) M_{a \vert x} \ \forall \ b,y, 
\end{align}
where $\mathcal{M}$ can be used to simulate~\cite{Guerini2017} the assemblage $\mathcal{M}'$ via the conditional probabilities $p(x \vert y)$ and $q(b \vert y,x,a)$ for all $y$, respectively for all $y,x,a$. 
Using the classical simulations, one also obtains the possible probabilities $q(y)$ to perform setting $y$ via $p(x) = \sum_y q(y) p(x \vert y)$. That means, it holds \cite{Tendick2023}:
\begin{align}
\label{monotone_simulation}
\Idiamond(\mathcal{M}^\mathbf{p}) \geq  \Idiamond(\xi(\mathcal{M}^\mathbf{p})^{\mathbf{q}}), 
\end{align}
for all measurement simulations $\xi$. Eq.~\eqref{monotone_simulation} follows ultimately from the fact that $\Idiamond(\mathcal{M}^\mathbf{p})$ is based on a norm and that it is written as a convex combination over the settings. \\
\indent Finally, since $\Idiamond(\mathcal{M}^\mathbf{p})$ is based on the diamond distance $\Ddiamond(\mathcal{M}^\mathbf{p}, \mathcal{N}^\mathbf{p}) \coloneqq \sum_x p(x) \Ddiamond(\Lambda_{\mathcal{M}_x}, \Lambda_{\mathcal{N}_x})$ between two weighted assemblages and the set $\mathrm{JM}$ of jointly measurable assemblage is convex, it is a convex function. Even more the distance $\Ddiamond(\mathcal{M}^\mathbf{p}, \mathcal{N}^\mathbf{p})$ fulfills the triangle inequality, i.e., 
\begin{align}
\Ddiamond(\mathcal{M}^\mathbf{p}, \mathcal{N}^\mathbf{p}) \leq \Ddiamond(\mathcal{M}^\mathbf{p}, \mathcal{L}^\mathbf{p}) + \Ddiamond(\mathcal{L}^\mathbf{p}, \mathcal{N}^\mathbf{p}),
\end{align}
for any weighted measurement assemblages $\mathcal{M}^\mathbf{p}, \mathcal{L}^\mathbf{p},$ and $\mathcal{N}^\mathbf{p}$. It therefore follows that
\begin{align}
\Idiamond(\mathcal{M}^\mathbf{p}) \leq  \Ddiamond(\mathcal{M}^\mathbf{p}, \mathcal{N}^{\#,\mathbf{p}}) \leq \Ddiamond(\mathcal{M}^\mathbf{p}, \mathcal{N}^\mathbf{p}) +   \Idiamond(\mathcal{N}^\mathbf{p}),
\end{align}
for any assemblages $\mathcal{M}$ and $\mathcal{N}$, where $\mathcal{N}^{\#} \in \mathrm{JM}$ is the closest jointly measurable assemblage with respect to $\mathcal{N}$. Note that the first inequality follows from the fact that $\mathcal{N}^{\#}$ is jointly measurable but not necessarily the closest jointly measurable assemblage to $\mathcal{M}$, i.e., $\mathcal{N}^{\#} \neq \mathcal{M}^{\#}$. \\
\indent To prove the tightness of our bounds in the main text, we rely on the \ac{SDP} formulation of $\Idiamond(\mathcal{M}^\mathbf{p})$, which besides its numerical uses allows us, in some instances, to determine 
the incompatibility of an assemblage analytically. 
In \cite{Tendick2023} it was shown that that $\Idiamond(\mathcal{M}^\mathbf{p})$ is equivalent to the optimal value of the \ac{SDP}:
\begin{align}
&\underline{\text{Primal problem (incompatibility):}} \label{Incomp_SDP_primal} \\
&\mathrm{given:} \ \mathcal{M}^\mathbf{p} \nonumber \\
&\underset{a_x, Z_x, G_{\lambda}}{\mathrm{minimize}} \sum_x p(x) a_x \nonumber \\
&\text{subject to:} \nonumber \\
 &a_x \mathds{1} - \mathrm{Tr}_1[Z_x] \geq 0 \ \forall \ a,x, \nonumber \\
 &Z_x \geq  \sum_a \vert a \rangle \langle a \vert \otimes (M_{a \vert x} - F_{a \vert x})^T \ \forall \ x, \nonumber \\
 &F_{a \vert x} = \sum_{\lambda} v(a \vert x, \lambda) G_{\lambda} \ \forall \ x,a, \ G_{\lambda} \geq 0 \ \forall \ \lambda, \sum_{\lambda} G_{\lambda} = \mathds{1}, \nonumber \\
&Z_x \geq 0, \ a_x \geq 0 \ \forall \ x, \nonumber 
\end{align}
where the $a_x$ are non-negative coefficients, the $Z_x$ are positive semidefinite matrices and the 
$G_{\lambda}$ are the \ac{POVM} effects of the parent \ac{POVM}. \acp{SDP} represent a special instance of convex optimization for which there exist off-the-shelf software~\cite{cvx_2,gb08_2,sdpt3_2,mosek_2} to efficiently solve them. Importantly, every \ac{SDP} comes with a \emph{dual formulation} that yields the same optimal value under some mild assumptions (see e.g.,~\cite{boyd_vandenberghe_2004}). This is indeed the case here \cite{Tendick2023}, i.e., $\Idiamond(\mathcal{M}^\mathbf{p})$ can also be understood as the optimal value of the \ac{SDP}:
\begin{align}
&\underline{\text{Dual problem (incompatibility):}} \label{Incomp_SDP_dual} \\
&\mathrm{given}: \ \mathcal{M}^\mathbf{p} \nonumber  \\
&\underset{C_{a \vert x}, \rho_x,L}{\mathrm{maximize}}  \ \ \ \sum_{a,x} p(x) \mathrm{Tr}[M_{a \vert x} C_{a \vert x}] - \mathrm{Tr}[L] \nonumber \\
&\text{subject to:} \nonumber \\
& L \geq \sum_{a,x} p(x) v(a \vert x, \lambda) C_{a \vert x} \ \forall \ \lambda, \nonumber \\
&0 \leq C_{a \vert x} \leq \rho_x \ \forall \ a,x, \ \rho_x \geq 0, \mathrm{Tr}[\rho_x] = 1 \ \forall \ x, \nonumber 
\end{align}
where the $C_{a \vert x}$, $\rho_x$, and $L$ are positive semidefinite matricies. Since the primal problem in Eq.~\eqref{Incomp_SDP_primal} corresponds to a minimization, every feasible point (i.e., any set of variables that fulfills all constraints) leads to an upper bound on $\Idiamond(\mathcal{M}^\mathbf{p})$. Similarly, every feasible solution of the dual in Eq.~\eqref{Incomp_SDP_dual} leads to a lower bound.

\section{Measurement splitting}
\label{SupplMeasurementSplitting}

In the main text, we argue that it is equivalent to consider the incompatibility of the assemblage $\mathcal{M}_{(1,2,1,3,2,3)}$ instead of $\mathcal{M}_{(1,2,3)}$, i.e., we use that $\Idiamond(\mathcal{M}_{(1,2,3)}) = \Idiamond(\mathcal{M}_{(1,2,1,3,2,3)})$ in order to derive the bound on the incompatibility gain in Eq.~(\textcolor{blue}{$10$}) in the main text.
Note that $\mathcal{M}_{(1,2,1,3,2,3)}$ is an assemblage in which each of the measurements $\mathcal{M}_1,\mathcal{M}_2$, and $\mathcal{M}_3$ occurs twice with probability $\tfrac{1}{6}$ each. On the other hand in $\mathcal{M}_{(1,2,3)}$ each of the measurements is used with a probability of $\tfrac{1}{3}$.
To show the equivalence $\Idiamond(\mathcal{M}_{(1,2,3)}) = \Idiamond(\mathcal{M}_{(1,2,1,3,2,3)})$, we actually show that $\Idiamond(\mathcal{M}_{(1,2,3)}) = \Idiamond(\mathcal{M}'_{(1,1,2,2,3,3)})$ and finally use that the set $\mathrm{JM}$ of jointly measurable measurements is closed under relabeling. We first show that $\mathcal{M}'_{(1,1,2,2,3,3)} = \xi(\mathcal{M}_{(1,2,3)})$ for a measurement simulation (see also Eq.~\eqref{MeasurementSimulation}) of the form
\begin{align}
M'_{b \vert y} = \sum_x p(x \vert y) \sum_a q(b \vert y,x,a) M_{a \vert x} \ \forall \ b,y, 
\end{align}
where we set $q(b \vert y,x,a) = \delta_{ba}$ for all $b,y,x,a$ with $\delta_{ba}$ being the Kronecker delta.
Furthermore, we use mixing probabilities $p(x \vert y)$ such that $p(x=1 \vert y =1) = p(x=1 \vert y =2) = 1$, $p(x=2 \vert y =3) = p(x=2 \vert y =4) = 1$, and $p(x=3 \vert y =5) = p(x=3 \vert y =6) = 1$ with all other probabilities set to zero. This is clearly a valid measurement simulation of $\mathcal{M}'_{(1,1,2,2,3,3)}$ using the measurements $\mathcal{M}_{(1,2,3)}$. Finally, notice that due to
$p(x) = \sum_y q(y) p(x \vert y)$, it holds 
\begin{align}
\label{Splitting_condition}
\dfrac{1}{3} = p(x=i) = q(y=2i-1)+q(y=2i), \ \text{for} \ i =1,2,3,
\end{align}
which is clearly fulfilled for $q(y) = \tfrac{1}{6} \ \forall y$. The above equation actually shows a more general statement, i.e., any probabilities $ p(y = 1) + p(y = 2)$ that sum to $\tfrac{1}{3}$ are allowed. This means, it is not necessary to split a \ac{POVM} into two equally likely \acp{POVM}, but one can introduce an additional bias. This bias will not change the results in a qualitative way, however, it can be used to fine-tune coefficients (multiplicative prefactors) such as changing the weights of the average in Eq. \textcolor{blue}{$(8)$} in the main text. The same holds for the other instances. \\
\indent To show the other direction, i.e., $\mathcal{M}_{(1,2,3)} = \xi(\mathcal{M}'_{(1,1,2,2,3,3)})$ we use again $q(b \vert y,x,a) = \delta_{ba}$ for all $b,y,x,a$. For the mixing probabilities, we set $p(x=1 \vert y =1) = p(x=2 \vert y =1) = \tfrac{1}{2}$, $p(x=3 \vert y =2) = p(x=4 \vert y =2) = \tfrac{1}{2}$, and  $p(x=5 \vert y =2) = p(x=6 \vert y =3) = \tfrac{1}{2}$ with all other probabilities set to zero. Again, it straightforward to check that this a valid measurement simulation. From the equivalence
\begin{align}
p(x = 1) = \dfrac{1}{6} = \sum_y q(y) p(1 \vert y) = q(y = 1) \dfrac{1}{2},
\end{align}
it follows directly that $q(y = 1) = \tfrac{1}{3}$ and similarly for the other cases. Now, since  $\mathcal{M}'_{(1,1,2,2,3,3)} = \xi(\mathcal{M}_{(1,2,3)})$ and $\mathcal{M}_{(1,2,3)} = \xi(\mathcal{M}'_{(1,1,2,2,3,3)})$, it holds that $\Idiamond(\mathcal{M}_{(1,2,3)}) = \Idiamond(\mathcal{M}'_{(1,1,2,2,3,3)})$. Analogously follows the measurement splitting with more measurements. \\
\indent Let us note here, that measurement simulations can also be used to show that the incompatibility of the parent \acp{POVM} of different subsets of jointly measurable assemblages is an upper bound on the incompatibility of these assemblages. More formally, let $\mathcal{M}_{(1,2,3)}$ be an assemblage and let 
$\mathcal{N}=\mathcal{M}^{\#}_{(1,2)} \mdoubleplus \mathcal{M}^{\#}_{(1,3)} \mdoubleplus \mathcal{M}^{\#}_{(2,3)}$ be the assemblage that contains the closest jointly measurable assemblages for the three subsets. Furthermore, let $\mathcal{G} = G(\mathcal{M}^{\#}_{(1,2)}) \mdoubleplus G(\mathcal{M}^{\#}_{(1,3)}) \mdoubleplus G(\mathcal{M}^{\#}_{(2,3)})$ be the assemblage that contains the parent \acp{POVM} of the respective subsets. With the above methods (and by the definition of the parent \ac{POVM} in Eq.~\eqref{Parent_POVM}) it can be seen that there exists a measurement simulation $\xi$ such that $\xi(\mathcal{G}) = \mathcal{N}$, which directly implies that $\Idiamond(\mathcal{N}) \leq \Idiamond(\mathcal{G})$ holds.

\section{Lower bounds}

Here, we prove the lower bounds stated in Eq.~(\textcolor{blue}{$13$}) and Eq.~(\textcolor{blue}{$16$}) in the main text. Remember, we consider the case in which $p(x) = \tfrac{1}{3}$, i.e., the input probabilities are uniformly distributed. Let us start by showing that 
\begin{align}
\Idiamond(\mathcal{M}_{(1,2,3)}) \geq \dfrac{1}{3}[\Idiamond(\mathcal{M}_{(1,2)})+\Idiamond(\mathcal{M}_{(1,3)}) +\Idiamond(\mathcal{M}_{(2,3)})],
\end{align}
holds. We start by using that $\Idiamond(\mathcal{M}_{(1,2,3)}) = \Idiamond(\mathcal{M}_{(1,2,1,3,2,3)})$. Now, the closest jointly measurable assemblage $\mathcal{M}^{\#}_{(1,2,1,3,2,3)}$ with respect to 
$\mathcal{M}_{(1,2,1,3,2,3)}$ allows us to rewrite $\Idiamond(\mathcal{M}_{(1,2,1,3,2,3)})$ such that
\begin{align}
\Idiamond(\mathcal{M}_{(1,2,1,3,2,3)}) = \Ddiamond(\mathcal{M}_{(1,2,1,3,2,3)},\mathcal{M}^{\#}_{(1,2,1,3,2,3)}).    
\end{align}
Now, concerning the measurement pairs $(1,2),(1,3),$ and $(2,3)$ the subsets of $\mathcal{M}^{\#}_{(1,2,1,3,2,3)}$ are jointly measurable by definition but not necessarily optimal for the respective subsets of $\mathcal{M}_{(1,2,1,3,2,3)}$. Using that the distance $\Ddiamond(\mathcal{M}_{(1,2,1,3,2,3)},\mathcal{M}^{\#}_{(1,2,1,3,2,3)})$ is a convex combination over the individual settings, it follows that 
\begin{align}
\Idiamond(\mathcal{M}_{(1,2,3)}) = \Idiamond(\mathcal{M}_{(1,2,1,3,2,3)}) \geq \dfrac{1}{3}[\Idiamond(\mathcal{M}_{(1,2)})+\Idiamond(\mathcal{M}_{(1,3)}) +\Idiamond(\mathcal{M}_{(2,3)})].  
\end{align}
To show the second lower bound, i.e., 
\begin{align}
\Idiamond(\mathcal{M}_{(1,2,3)}) \geq \tfrac{2}{3}\Idiamond(\mathcal{M}_{(1,2)}), 
\end{align}
it is enough to notice that leaving out the contribution of the setting $x=3$ can only lead to lower values than $\Idiamond(\mathcal{M}_{(1,2,3)})$. Finally, we use again that the remaining measurements (for the settings $x=1,2$) from the closest jointly measurable assemblage $\mathcal{M}^{\#}_{(1,2,3)}$ do not need to be optimal. 

\section{Steering and nonlocality}

Here, we show that our methods can directly be applied to quantum steering and Bell nonlocality. We start by considering steering. Let $\Vec{\sigma}_{(1,2,\cdots,m)} = (\sigma_1,\sigma_2,\cdots,\sigma_m)$ with $\sigma_x = \lbrace \sigma_{a \vert x}\rbrace_a$ be the \emph{steering assemblage} that Alice prepares for Bob by performing the measurements from a measurement assemblage $\mathcal{M}_{(1,2,\cdots,m)}$ on a shared state $\rho$ such that $\sigma_{a \vert x} = \mathrm{Tr}_A[(M_{a \vert x} \otimes \mathds{1})\rho]$. The \emph{consistent steering distance}~\cite{PhysRevA.97.022338} given by
\begin{align}
\label{steering_distance}
\mathrm{S}(\Vec{\sigma}) = \min\limits_{\Vec{\tau} \in \mathrm{CLHS}} \dfrac{1}{2}\sum\limits_{a,x} \dfrac{1}{m} \lVert \sigma_{a \vert x} - \tau_{a \vert x} \rVert_1, 
\end{align}
can be used to quantify the steerability of any steering assemblage $\mathrm{S}(\Vec{\sigma})$. Here, $\Vec{\tau} \in \mathrm{CLHS}$ denotes an assemblage that admits a \ac{LHS} and fulfills the consistency condition $\sum\limits_a \tau_{a\vert x} = \sum\limits_a \sigma_{a\vert x} = \rho_B = \mathrm{Tr}_A[\rho] \ \forall x.$ A \ac{LHS} for $\Vec{\tau}$ is given by 
\begin{align}
\tau_{a \vert x} = \sum_{\lambda} p(a \vert x, \lambda) \sigma_{\lambda},
\end{align}
where the $\sigma_\lambda$ are sub-normalized states and the $p(a \vert x, \lambda)$ resemble a classical post-processing, similarly to that in Eq.~\eqref{Parent_POVM} in the definition of jointly measurable assemblages. Note that we directly used here that the choice of the settings is uniformly distributed, i.e., $p(x) = \tfrac{1}{m}$. However, generally, we can use any distribution with $p(x) \neq 0 \ \forall x$, just like in the case for incompatibility. Note further that our following arguments are independent, as it was also the case for the incompatibility,  of the number of outcomes $a$ in the steering assemblage $\Vec{\sigma}.$ \\
\indent Now, since $\mathrm{S}(\Vec{\sigma})$ is based on a distance (the trace distance) we can directly derive the steering analog to the incompatibility bounds in the main text. In fact, our method relies only on the metric properties of the respective quantifiers, the fact they are written as a convex combination over the individual settings, and the general idea that a measurement can be split in two separate copies of itself. We make the following correspondence statements to our definitions for the incompatibility case: 
\begin{subequations}
\begin{align}
&\Vec{\sigma}_{(1,2,\cdots,m)} \longleftrightarrow \mathcal{M}_{(1,2, \cdots, m)}, \\
&\mathrm{S}(\Vec{\sigma}) \longleftrightarrow \Idiamond(\mathcal{M}), \\
&\Vec{\sigma}_{(1,2,\cdots,m)}^{\#}  \longleftrightarrow \mathcal{M}_{(1,2,\cdots,m)}^{\#}, \\
&\Vec{\sigma}_{(1,2,\cdots,m)}^{\# (1,2,\cdots,n)} \longleftrightarrow \mathcal{M}_{(1,2,\cdots,m)}^{\# (1,2,\cdots,n)}.
\end{align}
\end{subequations}
That is, $\Vec{\sigma}_{(1,2,\cdots,m)}^{\#}$ is the closest assemblage in the set $\mathrm{CLHS}$ to $\Vec{\sigma}_{(1,2,\cdots,m)}$ with respect to the distance
\begin{align}
\mathrm{D}_A(\Vec{\sigma}_{(1,2,\cdots,m)},\Vec{\sigma'}_{(1,2,\cdots,m)}) \coloneqq \sum_{a,x} \dfrac{1}{m} \lVert \sigma_{a \vert x} - \sigma'_{a \vert x} \rVert_1,
\end{align}
which induces the steering distance in Eq.~\eqref{steering_distance}. Furthermore, it holds
\begin{align}
\Vec{\sigma}_{(1,2,\cdots,m)}^{\# (1,2,\cdots,n)} \coloneqq 
\Vec{\sigma}_{(1,2,\cdots,n)}^{\#} \mdoubleplus \sigma_{n+1} \mdoubleplus \cdots \mdoubleplus \sigma_{m}.
\end{align}
\indent This implies, it holds that
\begin{align}
\mathrm{S}(\Vec{\sigma}_{(1,2,3)}) \leq \dfrac{1}{3}[\mathrm{S}(\Vec{\sigma}_{(1,2)})+\mathrm{S}(\Vec{\sigma}_{(1,3)})+\mathrm{S}(\Vec{\sigma}_{(2,3)})]+ \mathrm{S}(\Vec{\tau}),   
\end{align}
where $\Vec{\tau} =\Vec{\sigma}^{\#}_{(1,2)} \mdoubleplus \Vec{\sigma}^{\#}_{(1,3)} \mdoubleplus \Vec{\sigma}^{\#}_{(2,3)}$ is a state assemblage (with $m=6$ settings) that contains itself three assemblages (of two settings each) that are the closest consistent unsteerable assemblages to the respective subsets. Note that $\Vec{\tau}$ can be steerable in general. Note further that it is crucial to use a \emph{consistent} steering quantifier here, in order to avoid \emph{signaling} in the assemblage $\Vec{\tau}$. All the other bounds follow from here on directly. That is, it follows that
\begin{align}
\mathrm{S}(\Vec{\sigma}_{(1,2,3)}) &\geq \dfrac{1}{3}[\mathrm{S}(\Vec{\sigma}_{(1,2)})+\mathrm{S}(\Vec{\sigma}_{(1,3)})+\mathrm{S}(\Vec{\sigma}_{(2,3)})], \\
\mathrm{S}(\Vec{\sigma}_{(1,2,3)}) &\geq \dfrac{2}{3} \mathrm{S}(\Vec{\sigma}_{(1,2)}). \nonumber 
\end{align}
Moreover, using the assemblage $\Vec{\sigma}^{\# (1,2)}_{(1,2,3)}$ it holds that 
\begin{align}
\mathrm{S}(\Vec{\sigma}_{(1,2,3)}) \leq \dfrac{2}{3} \mathrm{S}(\Vec{\sigma}_{(1,2)}) + \mathrm{S}(\Vec{\sigma}^{\# (1,2)}_{(1,2,3)}).
\end{align}
\indent For nonlocality, very similar arguments can be made. However, we will see that additional constraints arise that distinguish nonlocality from steering and incompatibility. Let $\mathbf{q} = \lbrace q(ab \vert xy) \rbrace$ be a general probability distribution between two distant parties Alice and Bob. We consider the case where both, Alice and Bob, have two different measurement settings already available and Alice upgrades her measurement scheme with an additional third measurement. We denote the resulting distribution by $\mathbf{q}_{(1,2,3)}$. The nonlocality of a general distribution $\mathbf{q}$ can be quantified via the \emph{consistent} version of the classical trace distance quantifier introduced in~\cite{PhysRevA.97.022111}, which is given by
\begin{align}
\mathrm{N}(\mathbf{q}) = \dfrac{1}{2} \min\limits_{\mathbf{t} \in \mathrm{CLHV}} \sum\limits_{a,b,x,y} \dfrac{1}{m_A m_B} \lvert q(a,b \vert x,y) - t(a,b \vert x,y) \rvert.   \label{nonlocality_distance}
\end{align}
Here, we denote by $\mathrm{CLHV}$ the set of consistent \acp{LHV}, i.e., the set of those local distributions $\mathbf{t} \in \mathrm{LHV}$ that fulfill $\sum_a t(a,b \vert x,y) = t(b \vert y) = q(b \vert y) = \sum_a q(a,b \vert x,y) \ \forall \ b,y,x $ and similarly $\sum_b t(a,b \vert x,y) = t(a \vert x) = q(a \vert x) = \sum_b q(a,b \vert x,y) \ \forall \ a,y,x $. The (Bell) locality condition is expressed in terms of the \ac{LHV}: 
\begin{align}
t(a,b \vert x,y)  = \sum\limits_{\lambda} \pi(\lambda) p_A(a \vert x, \lambda) p_B(b \vert y, \lambda) \ \forall a,b,x,y, \label{LHV_model}
\end{align}
for the distribution $\mathbf{t}$. Finally, we denote by $m_A$ the number of measurement settings of Alice and by $m_B$ those of Bob, which we set to $m_B = 2$ here. Once again, we restrict our discussion to the case where the input probabilities $p(x,y) = p(x)p(y) = \tfrac{1}{m_A} \tfrac{1}{m_B}$ are uniformly distributed. \\
\indent Since $\mathrm{N}(\mathbf{q})$ relies on a distance that is written as a convex combination over the individual settings, we can use the triangle inequality together with the measurement splitting method. We make the following correspondence statements to our definitions in the incompatibility case:
\begin{subequations}
\begin{align}
&\mathbf{q}_{(1,2,\cdots,m_A)} \longleftrightarrow \mathcal{M}_{(1,2, \cdots, m)}, \\
&\mathrm{N}(\mathbf{q}) \longleftrightarrow \Idiamond(\mathcal{M}), \\
&\mathbf{q}^{\#}_{(1,2,\cdots,m_A)}  \longleftrightarrow \mathcal{M}_{(1,2,\cdots,m)}^{\#}, \\
&\mathbf{q}_{(1,2,\cdots,m_A)}^{\# (1,2,\cdots,n_A)} \longleftrightarrow \mathcal{M}_{(1,2,\cdots,m)}^{\# (1,2,\cdots,n)}.
\end{align}
\end{subequations}
That is, $\mathbf{q}^{\#}_{(1,2,\cdots,m_A)}$ is the closest consistent and local distribution to $\mathbf{q}_{(1,2,\cdots,m_A)}$ with respect the the classical trace distance ($\ell_1$ distance) that induces the nonlocality distance in Eq.~\eqref{nonlocality_distance}.
Furthermore, $\mathbf{q}_{(1,2,\cdots,m_A)}^{\# (1,2,\cdots,n_A)} = \mathbf{q}_{(1,2,\cdots,n_A)}^{\#} \mdoubleplus \mathbf{q}_{n_A+1} \mdoubleplus \cdots \mdoubleplus \mathbf{q}_{m_A}$, where we treat the probability vector that describes a distribution $\mathbf{q}_{(1,2,\cdots,m_A)}$ as ordered list. 
We would like to emphasize that the indices $(1,2,\cdots,m_A)$ refer to the measurements of Alice, and Bob's number of measurements remains fixed here. \\
\indent These correspondence relations imply that it is possible to obtain the bounds
\begin{align}
\dfrac{2}{3}\mathrm{N}(\mathbf{q}_{(1,2)}) \leq \mathrm{N}(\mathbf{q}_{(1,2,3)}) \leq \dfrac{2}{3}\mathrm{N}(\mathbf{q}_{(1,2)})+\mathrm{N}(\mathbf{q}^{\# (1,2)}_{(1,2,3)}),
\end{align}
Furthermore, we obtain the bounds
\begin{align}
\label{Nonlocality_bound}
\dfrac{1}{3}[\mathrm{N}(\mathbf{q}_{(1,2)})+\mathrm{N}(\mathbf{q}_{(1,3)})+\mathrm{N}(\mathbf{q}_{(2,3)})] \leq \mathrm{N}(\mathbf{q}_{(1,2,3)}) \leq  \dfrac{1}{3}[\mathrm{N}(\mathbf{q}_{(1,2)})+\mathrm{N}(\mathbf{q}_{(1,3)})+\mathrm{N}(\mathbf{q}_{(2,3)})] + \mathrm{N}(\mathbf{t}),
\end{align}
where $\mathbf{t}=\mathbf{q}^{\#}_{(1,2)} \mdoubleplus \mathbf{q}^{\#}_{(1,3)} \mdoubleplus \mathbf{q}^{\#}_{(2,3)}$ is a distribution (with $m_A = 6$ settings for Alice) which contains the closest local distributions with respect to the corresponding two-measurement subsets of Alice's measurement settings. \\
\indent Interestingly, the term $\dfrac{1}{3}[\mathrm{N}(\mathbf{q}_{(1,2)})+\mathrm{N}(\mathbf{q}_{(1,3)})+\mathrm{N}(\mathbf{q}_{(2,3)})]$ behaves differently from its steering and incompatibility counterpart. Namely, it is limited by the fact that $\mathrm{N}(\mathbf{q}_{(1,2)})$, $\mathrm{N}(\mathbf{q}_{(1,3)})$, and  $\mathrm{N}(\mathbf{q}_{(2,3)})$ cannot, in general, be maximal simultaneously. That is, contrary to incompatibility or steering, where all of the subset resources can be maximal at the same time. \\
\indent The reason for this is that there are not enough degrees of freedom for Alice to violate a given Bell inequality with different measurements, given that Bob keeps his settings fixed (besides the state that is also fixed). To exemplify this, we consider the scenario where both parties have two outcomes for each setting. In that case, the nonlocality of $\mathrm{N}(\mathbf{q}_{(1,2)})$, $\mathrm{N}(\mathbf{q}_{(1,3)})$, and $\mathrm{N}(\mathbf{q}_{(2,3)})$ is directly linked to the amount of violation of the \ac{CHSH} inequality~\cite{PhysRevLett.23.880}, as it was shown in~\cite{PhysRevA.97.022111}. However, the \ac{CHSH} inequality requires very specific combinations measurements to get maximal violation. Indeed, consider the three corresponding versions of the \ac{CHSH} inequality:
\begin{align}
\label{CHSH_inequalities}
\mathrm{CHSH}_{(1,2)} \coloneqq \langle A_1 \otimes B_1 \rangle + \langle A_1 \otimes B_2 \rangle + \langle A_2 \otimes B_1 \rangle - \langle A_2 \otimes B_2 \rangle \leq 2, \\
\mathrm{CHSH}_{(1,3)} \coloneqq \langle A_1 \otimes B_1 \rangle + \langle A_1 \otimes B_2 \rangle + \langle A_3 \otimes B_1 \rangle - \langle A_3 \otimes B_2 \rangle \leq 2, \nonumber \\
\mathrm{CHSH}_{(2,3)} \coloneqq \langle A_2 \otimes B_1 \rangle + \langle A_2 \otimes B_2 \rangle + \langle A_3 \otimes B_1 \rangle - \langle A_3 \otimes B_2 \rangle \leq 2, \nonumber
\end{align}
and their average
\begin{align}
\label{CHSH_123}
\mathrm{CHSH}_{(1,2,3)} \coloneqq \dfrac{1}{3}( \mathrm{CHSH}_{(1,2)}+\mathrm{CHSH}_{(1,3)}+\mathrm{CHSH}_{(2,3)})  \leq 2.
\end{align}
The inequality $\mathrm{CHSH}_{(1,2,3)} \leq 2$ can also be rewritten as
\begin{align}
\dfrac{2[\langle A_1 \otimes B_1 \rangle+\langle A_1 \otimes B_2 \rangle+\langle A_3 \otimes B_1 \rangle-\langle A_3 \otimes B_2 \rangle] + 2 \langle A_2 \otimes B_1 \rangle}{3} \leq 2,
\end{align}
which directly implies that the Tsirelson bound \cite{Cirelson1980}, i.e., the quantum bound of $\mathrm{CHSH}_{(1,2,3)}$ is given by $Q = \tfrac{4\sqrt{2}+2}{3} < 2\sqrt{2}$, i.e., quantum mechanics cannot reach the value $2\sqrt{2}$ that would correspond to all three contributions of Alice to be maximal simultaneously. The same is true for no-signaling theories, where the bound is given by $\mathrm{NS} = \tfrac{10}{3}$. Further, one needs to consider all possible combinations of different versions of the \ac{CHSH} inequalities in Eq.~\eqref{CHSH_inequalities}. That is, one needs to consider all $8$ symmetries of the \ac{CHSH} inequality, corresponding to the $8$ \ac{CHSH} facets of the local polytope. However, going through all the combinations shows that there is no combination which allows for a higher combined \ac{CHSH} value than $\mathrm{CHSH}_{(1,2,3)}$ in Eq.~\eqref{CHSH_123}.\\
\indent We want to emphasize again that such additional restrictions are not prevalent for the incompatibility and steering quantifiers analog of Eq.~\eqref{Nonlocality_bound}, which shows a clear separation of nonlocality to the other resources. Since the term $\dfrac{1}{3}[\mathrm{N}(\mathbf{q}_{(1,2)})+\mathrm{N}(\mathbf{q}_{(1,3)})+\mathrm{N}(\mathbf{q}_{(2,3)})]$ is also used in upper bounding the nonlocality  $\mathrm{N}(\mathbf{q}_{(1,2,3)})$, this could be a promising path to understanding why additional settings do not seem to increase the resource of nonlocality~\cite{Arajo2020,PhysRevA.97.022111}, in strict contrast to the resources of incompatibility and steerability. We expect that the same is true for more than two outcomes, however, more research in this direction is necessary.

\section{Generalizations}

In this section, we will generalize our framework from the main text in two directions. First, we discuss the scenario for more measurements i.e, $m > 3.$ Then, we will discuss the case in which the assemblage $\mathcal{M}^{\mathbf{p}} = (\mathcal{M},\mathbf{p})$ is weighted by a general probability distribution $\mathbf{p}$, instead of a uniform one. \\
\indent Using the methods from the main text and from Section~\ref{SupplMeasurementSplitting}, general bounds can be derived. We demonstrate this in the following for the assemblage $\mathcal{M}_{(1,2,3,4)}$ of $m=4$ uniformly distributed measurements. Further generalizations follow straightforwardly then. Let $\mathcal{M}^{\# (1,2,3)}_{(1,2,3,4)}$ be the closest assemblage with respect to the first three measurements of $\mathcal{M}_{(1,2,3,4)}$. Using the triangle inequality we get
\begin{align}
\Idiamond(\mathcal{M}_{(1,2,3,4)}) \leq \Ddiamond(\mathcal{M}_{(1,2,3,4)},\mathcal{M}^{\# (1,2,3)}_{(1,2,3,4)}) + \Idiamond(\mathcal{M}^{\# (1,2,3)}_{(1,2,3,4)}) = \dfrac{3}{4} \Idiamond(\mathcal{M}_{(1,2,3)}) + \Idiamond(\mathcal{M}^{\# (1,2,3)}_{(1,2,3,4)}),
\end{align}
as a direct generalization of Eq.~(\textcolor{blue}{$15$}) in the main text. \\
\indent In general, let $C_0 = \lbrace 1,2,\cdots,m \rbrace$ be the set of of all possible measurements from an assemblage $\mathcal{M}_{(1,2,\cdots,m)}$. Furthermore, let $C \in C_0$ be any non-empty subset of $C_0$ with cardinality $\lvert C \rvert$. It follows that 
\begin{align}
\label{GeneralBoound_1}
\Idiamond(\mathcal{M}_{(1,2,\cdots,m)}) \leq \dfrac{\lvert C \vert}{m} \Idiamond(\mathcal{M}_C) + \Idiamond(\mathcal{M}^{\# C}_{(1,2,\cdots,m)}),
\end{align}
where $\lvert C \vert$ is the number of measurements contained in the subset $C \in C_0$. Since Eq.~\eqref{GeneralBoound_1} holds for any subset $C$, we can conclude that 
\begin{align}
\label{GeneralBoound_2}
\Idiamond(\mathcal{M}_{(1,2,\cdots,m)}) \leq \min_{C \in C_0} \Big[\dfrac{\lvert C \vert}{m} \Idiamond(\mathcal{M}_C) + \Idiamond(\mathcal{M}^{\# C}_{(1,2,\cdots,m)})\Big],
\end{align}
which in particular includes the optimization over all 
$n$ measurement subsets. Note the upper bound trivially results in an equality in the case that $\lvert C \rvert = 1$, i.e., for practical purposes, one might exclude these cases from the minimization. \\ 
However, we can generalize our framework even more. Denote by $\lbrace C_i \rbrace$ a set of disjoint subsets of $C_0$ such that $\cup_i C_i = C_0$. It can directly be concluded that
\begin{align}
\label{GeneralBoound_3}
\Idiamond(\mathcal{M}_{(1,2,\cdots,m)}) \leq \sum_i \dfrac{\lvert C_i \rvert}{m} \Idiamond(\mathcal{M}_{C_i}) + \Idiamond(\mathcal{M}^{\#}_{C_1} \mdoubleplus \mathcal{M}^{\#}_{C_2} \mdoubleplus \dots \mdoubleplus \mathcal{M}^{\#}_{C_n}),
\end{align}
where $\mathcal{M}^{\#}_{C_1} \mdoubleplus \mathcal{M}^{\#}_{C_2} \mdoubleplus \dots \mdoubleplus \mathcal{M}^{\#}_{C_n}$ is an assemblage that contains itself the closest jointly measurable assemblages for the $n$ respective subsets $C_i$. Again, it is possible to minimize Eq.~\eqref{GeneralBoound_3} over a particular choice of different subsets and, in particular, over all non-trivial sets of subsets. \\
\indent Besides the generalization of our bounds based solely on particular instances of the triangle inequality, we can also use the measurement splitting method in a more general setup. For instance, by splitting each measurement from $\mathcal{M}_{(1,2,3,4)}$ three times, we obtain the assemblage $\mathcal{M}_{(1,2,3,1,2,4,1,3,4,2,3,4)}$. This lets us conclude that it holds
\begin{align}
\Idiamond(\mathcal{M}_{(1,2,3,4)}) \leq \dfrac{1}{4}[\Idiamond(\mathcal{M}_{(1,2,3)})+\Idiamond(\mathcal{M}_{(1,2,4)})+\Idiamond(\mathcal{M}_{(1,3,4)})+\Idiamond(\mathcal{M}_{(2,3,4)})] + \Idiamond(\mathcal{N}),  
\end{align}
with $\mathcal{N} = \mathcal{M}^{\#}_{(1,2,3)} \mdoubleplus \mathcal{M}^{\#}_{(1,2,4)} \mdoubleplus \mathcal{M}^{\#}_{(1,3,4)} \mdoubleplus \mathcal{M}^{\#}_{(2,3,4)}$ as a direct generalization of Eq.~(\textcolor{blue}{$8$}) in the main text. This leads for the generalization of the incompatibility gain in Eq.~(\textcolor{blue}{$10$}) in the main text to 
\begin{align}
\label{Bound_Incompatibility_Gain_4}
\Delta\mathrm{I}_{(1,2,3) \rightarrow (1,2,3,4)} \leq  \Idiamond(\mathcal{N}) \leq \Idiamond(\mathcal{G}),   
\end{align}
where we assume $\Idiamond(\mathcal{M}_{(1,2,3)}) \geq \max \lbrace \Idiamond(\mathcal{M}_{(1,2,4)}),\Idiamond(\mathcal{M}_{(1,3,4)}),\Idiamond(\mathcal{M}_{(2,3,4)}) \rbrace$ analogous to the condition stated in Result~\textcolor{blue}{$1$} in the main text and $\mathcal{G}$ is the assemblage containing the parent \acp{POVM} of the corresponding subsets. Again, further generalizations of~\eqref{Bound_Incompatibility_Gain_4} for other scenarios can be derived by applying our methods. \\
\indent We show, in the following, that our results can be applied to any probability distribution $\mathbf{p}$ with which an assemblage $\mathcal{M}$ is weighted. Here, we focus on assemblages with $m=3$ measurements. Further generalizations follow directly from the above discussion. Let $\mathcal{M}^{\mathbf{p}} =(\mathcal{M},\mathbf{p})$ be a general weighted measurement assemblage. Using the triangle inequality, it holds
\begin{align}
\Idiamond(\mathcal{M}_{(1,2,3)}^{\mathbf{p}}) \leq \Ddiamond(\mathcal{M}_{(1,2,3)}^{\mathbf{p}}, \mathcal{N}_{(1,2,3)}^{\mathbf{p}})+ \Idiamond(\mathcal{N}_{(1,2,3)}^{\mathbf{p}}),
\end{align}
for any assemblage $\mathcal{N}_{(1,2,3)}$. By setting $ \mathcal{N} = \mathcal{M}^{\# (1,2)}_{(1,2,3)} \coloneqq \mathcal{M}^{\#}_{(1,2)} \mdoubleplus \mathcal{M}_3$, it follows that 
\begin{align}
\Ddiamond(\mathcal{M}_{(1,2,3)}^{\mathbf{p}}, \mathcal{N}_{(1,2,3)}^{\mathbf{p}}) = [p(1)+p(2)]  \Idiamond(\mathcal{M}_{(1,2)}^{\mathbf{q}}),
\end{align}
where $\mathbf{q} = (\tfrac{p(1)}{p(1)+p(2)},\tfrac{p(2)}{p(1)+p(2)})$ is the probability distribution weighting the assemblage $\mathcal{M}_{(1,2)}$. It is important to note here, that 
$\mathcal{M}^{\#}_{(1,2)}$ refers specifically to the closest assemblage to $\mathcal{M}_{(1,2)}$ with respect to the distribution $\mathbf{q}$. Note further that the particular instance of a uniform distribution can straightforwardly be recovered from here. This shows that $\Idiamond(\mathcal{M}_{(1,2,3)}^{\mathbf{p}})$ is upper bounded by the incompatibility of its subset $\mathcal{M}_{(1,2)}$ weighted by the likelihood of choosing a measurement from that subset, plus the incompatibility $\Idiamond(\mathcal{M}^{\# (1,2),\mathbf{p}}_{(1,2,3)})$. Similarly, if we want to use the measurement splitting method, we can chose any initial distribution $\mathbf{p}$ and proceed as usual to obtain bounds. As we noted in Section~\ref{SupplMeasurementSplitting} the method is not limited to split a measurement into two equally likely versions of itself. The only conditions that have to be satisfied are the conditions in the second equality of Eq.~\eqref{Splitting_condition}.

\section{Proofs regarding the incompatibility of mutually unbiased bases}

In this section, we present the proofs related to statements in the main text regarding the incompatibility of measurements based on \ac{MUB}~\cite{DURT2010}. Two orthonormal bases $\lbrace \vert v_a \rangle \rbrace_{0 \leq a \leq d-1}$ and $\lbrace \vert w_b \rangle \rbrace_{0 \leq b \leq d-1}$ are said to be \ac{MUB} if it holds that
\begin{align}
\vert \langle v_a \vert w_b \rangle \vert = \dfrac{1}{\sqrt{d}} \ \forall \ a,b. \label{Overlap_MUB}
\end{align}
The set of projections onto the orthonormal bases $\lbrace \vert v_a \rangle \rbrace_{0 \leq a \leq d-1}$ form the measurement $ \mathcal{M} = \lbrace M_a = \lvert v_a \rangle \langle v_a \rvert \rbrace$. Now, an \emph{\ac{MUB} measurement assemblage} \cite{Tendick2023} is a set of measurements where the condition~\eqref{Overlap_MUB} holds for any two projections from different bases. While it is generally unknown how many \ac{MUB} exist in a dimension $d$, it is known that for every $d \geq 2$, there exist at least $m=p^r+1$ \ac{MUB}, where $p^r$ is the smallest prime power factor of $d$ \cite{10.1007/978-3-540-24633-6_10}, and at most $m=d+1$ \ac{MUB}. In the case where $d$ is a prime-power there exist explicit constructions of \ac{MUB}~\cite{Wootters1989}, which are known to be operationally inequivalent~\cite{PhysRevLett.122.050402,Tendick2023}. The possibly most simple construction of a complete set of \ac{MUB}, i.e., $m=d+1$ bases, can be used whenever $d$ is a prime. In this case, we can use the Heisenberg-Weyl operators
\begin{align}
\hat{X} = \sum_{k=0}^{d-1} \lvert k+1 \rangle \langle k \rvert, \ \hat{Z} = \sum_{k=0}^{d-1} \omega^k \lvert k \rangle \langle k \rvert,    
\end{align}
for a specific construction. Here, $\lbrace \lvert k \rangle \rbrace_{0 \leq k \leq d-1}$ is the computational basis and $\omega = \exp{\big(\dfrac{2\pi i}{d}\big)}$ is a root of unity. In prime dimensions $d$, the eigenbases of the $d+1$ operators $\hat{X},\hat{Z},\hat{X} \hat{Z}, \hat{X} \hat{Z}^2, \cdots, \hat{X}\hat{Z}^{d-1}$ are mutually unbiased~\cite{bandyopadhyay2002new}. Most notably, for $m=2$, $m = d,$ and $m=d+1$ there exists an analytical expression for the incompatibility of the \ac{MUB} measurement assemblages obtained via this construction ~\cite{PhysRevLett.122.050402,Tendick2023}. For $d=2$, our \ac{MUB} measurement assemblage reduces to the projective measurements defined by the Pauli operators. 

\subsection{Tightness proof for Eq.~(\textcolor{blue}{$10$}) in the main text}
\label{TightnessProof_1}

\indent We start by proving that Eq.~(\textcolor{blue}{$10$}) in the main text is tight for a noisy \ac{MUB} measurement assemblage based on Pauli measurements, i.e., we show that 
\begin{align}
\Delta\mathrm{I}_{(1,2) \rightarrow (1,2,3)}(\eta) \coloneqq  \Idiamond(\mathcal{M}^{\eta}_{(1,2,3)})-\Idiamond(\mathcal{M}^{\eta}_{(1,2)}) =  \Idiamond(\mathcal{N}(\eta)), 
\end{align}
with $\mathcal{N}(\eta)= \mathcal{M}^{\# \eta}_{(1,2)} \mdoubleplus \mathcal{M}^{\# \eta}_{(1,3)} \mdoubleplus \mathcal{M}^{\# \eta}_{(2,3)}$ holds true for measurements of the form
\begin{align}
\label{NoisyPauliSuppl}
M^{\eta}_{a \vert x} = \eta \Pi_{a \vert x} + (1-\eta) \Tr[\Pi_{a \vert x}] \dfrac{\mathds{1}}{2}, 
\end{align}
where the $\Pi_{a \vert x} = M^{\eta = 1}_{a \vert x} $ are projectors defined via the eigenvectors of Pauli operators and $\eta$ defines the amount of noise in the measurements. \\
\indent We divide our proof into three different parameter regimes. Let $\eta^*_2$ and $\eta^*_3$ be the \emph{white-noise robustness} of $\mathcal{M}_{(1,2)}$, respectively $\mathcal{M}_{(1,2,3)}$, i.e., the maximal $\eta$ where the noisy assemblages are still jointly measurable. We consider the regimes $1): \ \eta \leq \eta^*_3 \leq \eta^*_2$, \ $2): \ \eta^*_3 < \eta \leq \eta^*_2$, and $3): \ \eta^*_3 \leq \eta^*_2 < \eta$ corresponding to the three regimes in Figure~\textcolor{blue}{$2$} in the main text. \\
\indent Note that regime $1): \ \eta \leq \eta^*_3$ leads trivially to
\begin{align}
\Delta\mathrm{I}_{(1,2) \rightarrow (1,2,3)}(\eta) = \Idiamond(\mathcal{N}(\eta)) = 0.   
\end{align}
For the second regime, i.e., $\eta^*_3 < \eta \leq \eta^*_2$ it follows directly that 
\begin{align}
\Delta\mathrm{I}_{(1,2) \rightarrow (1,2,3)}(\eta) = \Idiamond(\mathcal{M}^{\eta}_{(1,2,3)}) = \Idiamond(\mathcal{N}(\eta)).
\end{align}
The first equality follows from the fact that $\Idiamond(\mathcal{M}^{\eta}_{(1,2)}) = 0$ by definition. The second equality follows from the fact that $\mathcal{M}^{\# \eta}_{(s,t)} = \mathcal{M}^{\eta}_{(s,t)}$ for any $s,t \in \lbrace 1,2,3 \rbrace$ such that $s \neq t$, since the subset $\mathcal{M}^{\eta}_{(s,t)}$ is jointly measurable by definition. Now, due to the reverse direction of the measurement splitting method outlined in Section \ref{SupplMeasurementSplitting}, it holds $\Idiamond(\mathcal{M}^{\eta}_{(1,2,3)}) = \Idiamond(\mathcal{N}(\eta))$. That means, the only non-trivial case is regime $3): \ \eta^*_3 \leq \eta^*_2 < \eta$. \\
\indent Our proof for this regime relies on solving the \acp{SDP} in Eq.~\eqref{Incomp_SDP_primal} and Eq.~\eqref{Incomp_SDP_dual} analytically. Starting from the dual:
\begin{align}
&\underline{\text{Dual problem (incompatibility):}}  \\
&\mathrm{given}: \ \mathcal{M}^{\eta}, \mathbf{p} \nonumber  \\
&\underset{C_{a \vert x}, \rho_x,L}{\mathrm{maximize}}  \ \ \ \sum_{a,x} p(x) \mathrm{Tr}[M^{\eta}_{a \vert x} C_{a \vert x}] - \mathrm{Tr}[L] \nonumber \\
&\text{subject to:} \nonumber \\
& L \geq \sum_{a,x} p(x) v(a \vert x, \lambda) C_{a \vert x} \ \forall \ \lambda, \nonumber \\
&0 \leq C_{a \vert x} \leq \rho_x \ \forall \ a,x, \ \rho_x \geq 0, \mathrm{Tr}[\rho_x] = 1 \ \forall \ x, \nonumber 
\end{align}
we choose the specific instance where $C_{a \vert x} = \tfrac{\Pi_{a \vert x}}{2}$, $L = l \mathds{1}$, and $\rho_x = \sum_a C_{a \vert x} = \tfrac{\mathds{1}}{2}$ for some appropriately chosen scalar-variable $l$. For a qubit assemblage $\mathcal{M}$ with \ac{POVM} effects of the form
\begin{align}
M^{\eta}_{a \vert x} = \eta \Pi_{a \vert x} + (1-\eta) \Tr[\Pi_{a \vert x}] \dfrac{\mathds{1}}{2} = \eta \Pi_{a \vert x} + (1-\eta) \dfrac{\mathds{1}}{2},     
\end{align}
this evaluates to the lower bound $\Idiamond(\mathcal{M}^{\eta}) \geq \eta +\tfrac{(1-\eta)}{2} - \tfrac{T}{m}$, where $T \coloneqq \lVert \sum_{a,x} v^*(a \vert x,\lambda) M_{a \vert x} \rVert_{\infty}$ and $\lbrace v^*(a \vert x,\lambda) \rbrace_{a,x}$ is the deterministic strategy maximizing the norm. Note that this bound results from choosing $l = \tfrac{T}{2m}$, which can be shown to be always a valid choice \cite{Tendick2023}. \\ 
\indent For (noise-free, i.e., $\eta = 1$) \ac{MUB} measurement assemblages it was proven in~\cite{PhysRevLett.122.050402} that whenever $m=2$, $m=d$, or $m=d+1$, it holds that
\begin{align}
\eta^*_m = \dfrac{dT-m}{dm-m}.    
\end{align}
This lets us conclude (for the qubit case, i.e., d = 2) that
\begin{align}
\Idiamond(\mathcal{M}^{\eta}) &\geq  \eta +\dfrac{(1-\eta)}{2} - \dfrac{\eta^*_m+1}{2} \\
&= \dfrac{1}{2}(\eta-\eta^*_m) \nonumber.
\end{align}
For the upper bound of $\Idiamond(\mathcal{M}^{\eta})$ we invoke the primal \ac{SDP}: 
\begin{align}
&\underline{\text{Primal problem (incompatibility):}}  \label{PrimalProblemRep}\\
&\mathrm{given:} \ \mathcal{M}^{\eta}, \mathbf{p} \nonumber \\
&\underset{Z_x, G_{\lambda}}{\mathrm{minimize}} \sum_x p(x) \lVert \mathrm{Tr}_1[Z_x] \rVert_{\infty} \nonumber \\
&\text{subject to:} \nonumber \\
&Z_x \geq  \sum_a \vert a \rangle \langle a \vert \otimes (M^{\eta}_{a \vert x} - F_{a \vert x})^T \ \forall \ x, \nonumber \\
 &F_{a \vert x} = \sum_{\lambda} v(a \vert x, \lambda) G_{\lambda} \ \forall \ x,a, \ G_{\lambda} \geq 0 \ \forall \ \lambda, \sum_{\lambda} G_{\lambda} = \mathds{1}, \nonumber \\
&Z_x \geq 0,  \ \forall \ x, \nonumber 
\end{align}
where we have explicitly replaced the constraints involving the variables $a_x$ in the \ac{SDP} in Eq.~\eqref{Incomp_SDP_primal} by using the spectral norm (largest singular value). By choosing $F_{a \vert x} = \eta^*_m \Pi_{a \vert x} +(1-\eta^*_m)\tfrac{\mathds{1}}{2}$ and $Z_x = \tfrac{1}{2}(\eta-\eta^*_m) \sum_a \lvert a \rangle \langle a \rvert \otimes \Pi_{a \vert x}^T$ for $\eta \geq \eta^*_m$ all constraints can directly be verified to hold. Therefore, we obtain the upper bound
\begin{align}
\Idiamond(\mathcal{M}^{\eta}) \leq \dfrac{1}{2}(\eta-\eta^*_m).
\end{align}
That implies $\Idiamond(\mathcal{M}^{\eta}) = \dfrac{1}{2}(\eta-\eta^*_m)$ for any assemblage involving $m=2$ or $m=3$ noisy \ac{MUB} measurement assemblages with $\eta \geq \eta^*_m$ in $d=2$. 
Therefore, the incompatibility gain $\Delta\mathrm{I}_{(1,2) \rightarrow (1,2,3)}(\eta) \coloneqq \Idiamond(\mathcal{M}^{\eta}_{(1,2,3)})-\Idiamond(\mathcal{M}^{\eta}_{(1,2)})$ evaluates to
\begin{align}
\Delta\mathrm{I}_{(1,2) \rightarrow (1,2,3)}(\eta) &= 
\dfrac{1}{2}[(\eta - \eta)+(\eta^*_2-\eta^*_3)]. \\
&= \dfrac{1}{2}[(\eta^*_2-\eta^*_3)]. \nonumber
\end{align}
Note that the gain is constant in this regime, as it is also evident from Figure~\textcolor{blue}{$2$} in the main text.
Now, to finish the proof, we have to show that $\Idiamond(\mathcal{N}(\eta))$ has the same incompatibility. However, this follows almost directly, since $\mathcal{N}(\eta)= \mathcal{M}^{\# \eta}_{(1,2)} \mdoubleplus \mathcal{M}^{\# \eta}_{(1,3)} \mdoubleplus \mathcal{M}^{\# \eta}_{(2,3)}$ contains the closest jointly measurable assemblages with respect to the subsets. As it is known from~\cite{PhysRevLett.122.050402} and~\cite{Tendick2023} (and we confirmed it with the above calculation) all of these subsets are again just noisy versions of \ac{MUB} measurement assemblages, with the same noise contained in every subset. From the reverse direction of the measurement splitting method, it follows that
\begin{align}
\Idiamond(\mathcal{N}(\eta)) = \Idiamond(\mathcal{M}^{\eta}_{(1,2,3)}) 
\end{align}
for $\eta = \eta^*_2 = \tfrac{1}{\sqrt{2}}$. Therefore, it follows that $\Idiamond(\mathcal{N}(\eta)) = \dfrac{1}{2}[(\eta^*_2-\eta^*_3)]$ for $ \eta \geq \eta^*_2 $ which concludes the proof. \\

\subsection{Tightness proof for Eq.~(\textcolor{blue}{$15$}) in the main text}

\begin{figure}
\includegraphics[scale=0.5]{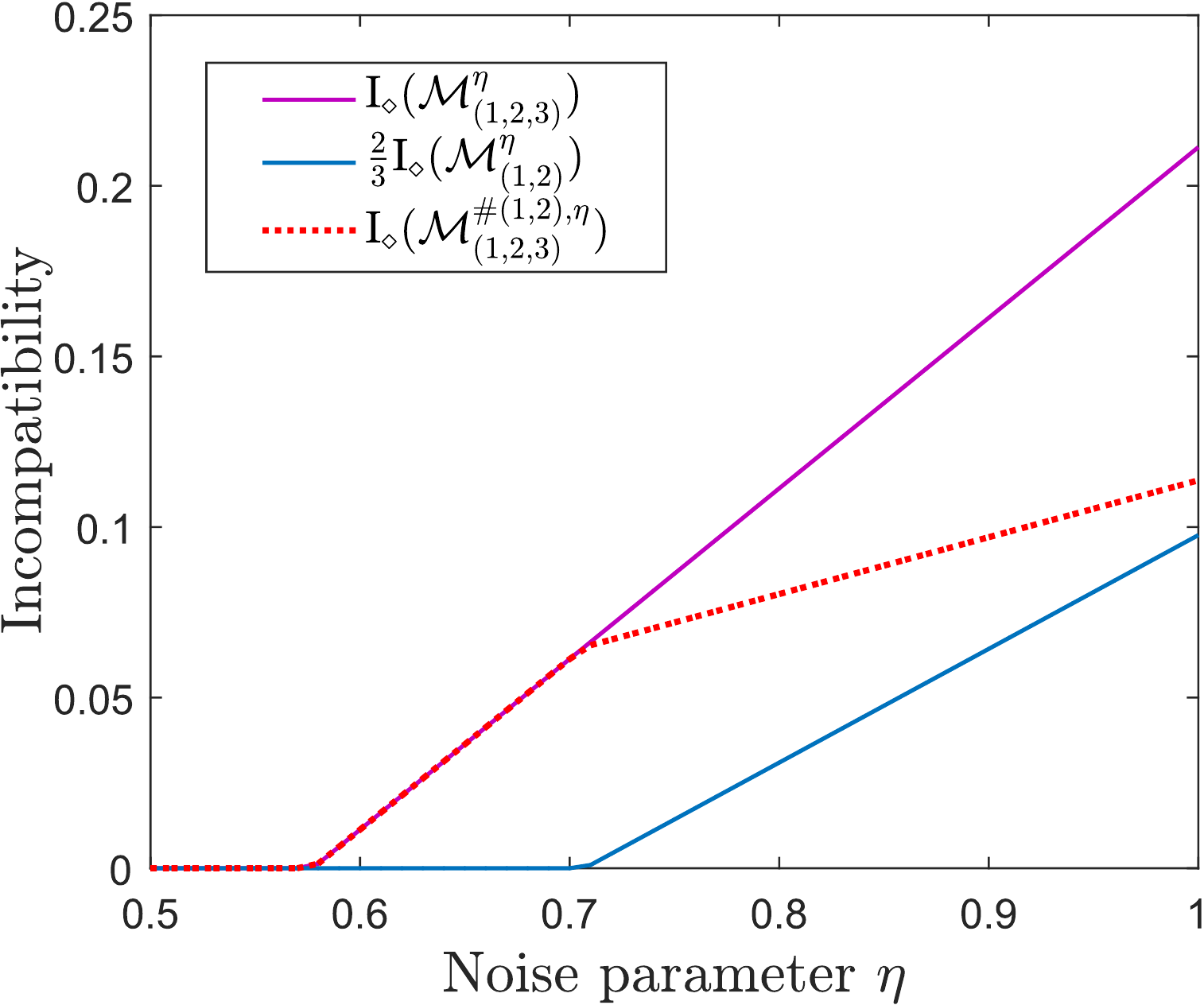}
 \caption{Incompatibility bound from Eq.~(\textcolor{blue}{$15$}) in the main text for measurements corresponding to the three Pauli measurements. The different contributions (depicted by the dashed red and the blue line) result together in the incompatibility $\Idiamond(\mathcal{M}^{\eta}_{(1,2,3)})$. The two discontinuities of  $\Idiamond(\mathcal{M}^{\# (1,2), \eta}_{(1,2,3)})$ (dashed red line) indicate the points where $\mathcal{M}^{\eta}_{(1,2,3)}$ becomes compatible, respectively pairwise compatible.}
  \label{figure_Incompatiblity_simple_upper_bound}
\end{figure}

Here, we show that Eq.~(\textcolor{blue}{$15$}) in the main text is tight for the case of noisy Pauli measurements (see Eq.~\eqref{NoisyPauliSuppl}). That is, we show that 
\begin{align}
\Idiamond(\mathcal{M}^{\eta}_{(1,2,3)}) = \dfrac{2}{3} \Idiamond(\mathcal{M}^{\eta}_{(1,2)}) + 
\Idiamond(\mathcal{M}^{\# (1,2), \eta}_{(1,2,3)}),
\end{align}
holds for the assemblage $\mathcal{M}^{\eta}_{(1,2,3)}$ that contains noisy Pauli measurements. To give a better overview, we also plot the respective incompatibility contributions of $\Idiamond(\mathcal{M}^{\eta}_{(1,2,3)})$ in Figure \ref{figure_Incompatiblity_simple_upper_bound}.
The proof reduces to show the equality for the case $\eta > \eta^*_2$, as the other cases follow trivially from the discussions made in Section~\ref{TightnessProof_1}. We already evaluated the values of $\Idiamond(\mathcal{M}^{\eta}_{(1,2,3)})$ and $\Idiamond(\mathcal{M}^{\eta}_{(1,2)})$, i.e., we only have to show that 
\begin{align}
\Idiamond(\mathcal{M}^{\# (1,2), \eta}_{(1,2,3)}) = \Idiamond(\mathcal{M}^{\eta}_{(1,2,3)})-\dfrac{2}{3} \Idiamond(\mathcal{M}^{\eta}_{(1,2)}) &= \dfrac{1}{2}(\eta-\eta^*_3)-\dfrac{1}{3}(\eta-\eta^*_2) \\
&= \dfrac{1}{6} \eta + \dfrac{1}{3} \eta^*_2 - \dfrac{1}{2}\eta^*_3. \nonumber
\end{align}
Since we already know that $\dfrac{1}{6} \eta + \dfrac{1}{3} \eta^*_2 - \dfrac{1}{2}\eta^*_3 \leq \Idiamond(\mathcal{M}^{\# (1,2), \eta}_{(1,2,3)}),$ due to the general bound in Eq.~\eqref{GeneralBoound_1}, it is enough to show that $\Idiamond(\mathcal{M}^{\# (1,2), \eta}_{(1,2,3)}) \leq \dfrac{1}{6} \eta + \dfrac{1}{3} \eta^*_2 - \dfrac{1}{2}\eta^*_3 $ also holds true. \\
\indent We rely again on the primal problem in Eq.~\eqref{PrimalProblemRep} using the feasible point where $F_{a \vert x} = \eta^*_3 \Pi_{a \vert x} +(1-\eta^*_3)\tfrac{\mathds{1}}{2}$ 
and  $Z_x = \tfrac{1}{2}(\eta^*_2-\eta^*_3) \sum_a \lvert a \rangle \langle a \rvert \otimes \Pi_{a \vert x}^T$ for $x = 1,2$ and $Z_x = \tfrac{1}{2}(\eta-\eta^*_3) \sum_a \lvert a \rangle \langle a \rvert \otimes \Pi_{a \vert x}^T$ for $x = 3$. It can be checked again directly that this point is indeed feasible. Moreover, we obtain a primal objective value of
\begin{align}
 \sum_x \dfrac{1}{3} \lVert \mathrm{Tr}_1[Z_x] \rVert_{\infty}    = 2\dfrac{1}{3}\cdot \dfrac{1}{2}(\eta^*_2-\eta^*_3)+\dfrac{1}{6}(\eta-\eta^*_3) =  \dfrac{1}{6} \eta + \dfrac{1}{3} \eta^*_2 - \dfrac{1}{2}\eta^*_3,
\end{align}
which concludes the proof.

\subsection{Tightness proof for generalizations of Eq.~(\textcolor{blue}{$15$}) in the main text}

Here, we show that in the scenarios $m=2\rightarrow m'=d$, $m=2\rightarrow m'=d+1$, and $m=d\rightarrow m'=d+1$ there exists analog bounds to Eq.~(\textcolor{blue}{$15$}) in the main text that are tight for $d$-dimensional noisy \ac{MUB} measurement assemblages. As before, we only consider the non-trivial case here and in the following, i.e., the noisy regime where none of the incompatibilities vanish. Also, we refer to the noise-free measurements, i.e., the projectors on the \ac{MUB} by $\Pi_{a \vert x} = M_{a \vert x}^{\eta = 1}$. The corresponding bound (see Eq.~\eqref{GeneralBoound_1}) for the instance $2\rightarrow d$ reads
\begin{align}
\Idiamond(\mathcal{M}^{\eta}_{(1,2,\cdots,d)}) \leq \dfrac{2}{d} \Idiamond(\mathcal{M}^{\eta}_{(1,2)}) + 
\Idiamond(\mathcal{M}^{\# (1,2), \eta}_{(1,2,\cdots,d)}),
\end{align}
where we know that $\Idiamond(\mathcal{M}^{\eta}_{(1,2)})= (\tfrac{d-1}{d})(\eta-\eta^*_2)$ by generalizing the previous qubit result. Indeed, carefully checking the calculation for the $d=2$ case in the Section \ref{TightnessProof_1}, reveals the general (dimension dependant) prefactor for the incompatibility of two noisy \ac{MUB} measurements. \\
\indent Furthermore, using essentially the same feasible points as before (simply extended to the case of $m=d$ instead of $m=2$ measurements) we obtain that $\Idiamond(\mathcal{M}^{\eta}_{(1,2,\cdots,d)}) = (\tfrac{d-1}{d})(\eta-\eta^*_d)$. With that, we know that 
\begin{align}
\Idiamond(\mathcal{M}^{\# (1,2), \eta}_{(1,2,\cdots,d)}) \geq \Idiamond(\mathcal{M}^{\eta}_{(1,2,\cdots,d)})-\dfrac{2}{d} \Idiamond(\mathcal{M}^{\eta}_{(1,2)}) = \Big(\dfrac{d-1}{d}\Big)(\eta-\eta^*_d)-\dfrac{2}{d}\Big(\dfrac{d-1}{d}\Big)(\eta-\eta^*_2),    
\end{align}
which means, it remains to show that $\Idiamond(\mathcal{M}^{\# (1,2), \eta}_{(1,2,\cdots,d)}) \leq (\tfrac{d-1}{d})(\eta-\eta^*_d)-\tfrac{2}{d}(\tfrac{d-1}{d})(\eta-\eta^*_2)$ also holds. This can directly be verified by using the feasible point $F_{a \vert x} = \eta^*_d \Pi_{a \vert x} +(1-\eta^*_d)\tfrac{\mathds{1}}{d}$ and  $Z_x = (\tfrac{d-1}{d})(\eta^*_2-\eta^*_d) \sum_a \lvert a \rangle \langle a \rvert \otimes \Pi_{a \vert x}^T$ for $x=1,2,$ and $Z_x = (\tfrac{d-1}{d})(\eta-\eta^*_d) \sum_a \lvert a \rangle \langle a \rvert \otimes \Pi_{a \vert x}^T$ for $x=3, \cdots,d$. This concludes the proof. \\
\indent The corresponding bound for the $2\rightarrow d+1$ scenario (see Eq.~\eqref{GeneralBoound_1}) reads
\begin{align}
\Idiamond(\mathcal{M}^{\eta}_{(1,2,\cdots,d+1)}) \leq \dfrac{2}{d+1} \Idiamond(\mathcal{M}^{\eta}_{(1,2)}) + 
\Idiamond(\mathcal{M}^{\# (1,2), \eta}_{(1,2,\cdots,d+1)}),
\end{align}
with $\Idiamond(\mathcal{M}^{\eta}_{(1,2)}) = (\tfrac{d-1}{d})(\eta-\eta^*_2)$. Using the same feasible points as for the $m=2$ and $m=d$ case, it also follows that $\Idiamond(\mathcal{M}^{\eta}_{(1,2,\cdots,d+1)}) = (\tfrac{d-1}{d})(\eta-\eta^*_{d+1})$, i.e, to prove tightness, we have to show that 
\begin{align}
\label{Bound_2_d+1}
 \Idiamond(\mathcal{M}^{\# (1,2), \eta}_{(1,2,\cdots,d+1)})  \leq  \Big(\dfrac{d-1}{d}\Big)(\eta-\eta^*_{d+1})-\Big(\dfrac{2}{d+1}\Big)\Big(\dfrac{d-1}{d}\Big)(\eta-\eta^*_2),
\end{align}
holds true. Using the same construction as before, this can be checked directly. Namely, using the feasible point $F_{a \vert x} = \eta^*_{d+1} \Pi_{a \vert x} +(1-\eta^*_{d+1})\tfrac{\mathds{1}}{d}$ and  $Z_x = (\tfrac{d-1}{d})(\eta^*_2-\eta^*_{d+1}) \sum_a \lvert a \rangle \langle a \rvert \otimes \Pi_{a \vert x}^T$ for $x=1,2,$ and $Z_x = (\tfrac{d-1}{d})(\eta-\eta^*_{d+1}) \sum_a \lvert a \rangle \langle a \rvert \otimes \Pi_{a \vert x}^T$ for $x=3, \cdots,d+1$ it follows directly that Eq.~\eqref{Bound_2_d+1} is indeed true, which concludes the proof. \\
\indent In the case $d\rightarrow d+1$, the corresponding bound reads
\begin{align}
\Idiamond(\mathcal{M}^{\eta}_{(1,2,\cdots,d+1)}) \leq \dfrac{d}{d+1} \Idiamond(\mathcal{M}^{\eta}_{(1,2,\cdots,d)}) + 
\Idiamond(\mathcal{M}^{\# (1,2,\cdots,d), \eta}_{(1,2,\cdots,d+1)}),
\end{align}
with $\Idiamond(\mathcal{M}^{\eta}_{(1,2,\cdots,d+1)}) = (\tfrac{d-1}{d})(\eta-\eta^*_{d+1})$ and $\Idiamond(\mathcal{M}^{\eta}_{(1,2,\cdots,d)}) = (\tfrac{d-1}{d})(\eta-\eta^*_{d})$. That means we have to check that 
\begin{align}
\label{Bound_d_d+1}
 \Idiamond(\mathcal{M}^{\# (1,2,\cdots,d), \eta}_{(1,2,\cdots,d+1)})  \leq  \Big(\dfrac{d-1}{d}\Big)(\eta-\eta^*_{d+1})-\Big(\dfrac{d}{d+1}\Big)\Big(\dfrac{d-1}{d}\Big)(\eta-\eta^*_{d}),
\end{align}
is true. Using $F_{a \vert x} = \eta^*_{d+1} \Pi_{a \vert x} +(1-\eta^*_{d+1})\tfrac{\mathds{1}}{d}$ and  $Z_x = \tfrac{d-1}{d}(\eta^*_d-\eta^*_{d+1}) \sum_a \lvert a \rangle \langle a \rvert \otimes \Pi_{a \vert x}^T$ for $x=1,2,\cdots,d,$ and $Z_x = \tfrac{d-1}{d}(\eta-\eta^*_{d+1}) \sum_a \lvert a \rangle \langle a \rvert \otimes \Pi_{a \vert x}^T$ for $x=d+1$ this can be verified, just as in the above cases.

\subsection{Additional insights on Eq.~(\textcolor{blue}{$18$}) in the main text}

In this subsection, we give additional insights to Eq.~(\textcolor{blue}{$18$}) from the main text. That is, we analyse the incompatibility decomposition 
\begin{align}
\label{SupplIncompatiblityDecomposition}
\Idiamond(\mathcal{M}_{(1,2,3)}) &\leq \Idiamond^{\mathrm{gen}}(\mathcal{M}_{(1,2,3)}) + \Idiamond^{\mathrm{pair}}(\mathcal{M}_{(1,2,3)})+\Idiamond^{\mathrm{hol}}(\mathcal{M}_{(1,2,3)}),
\end{align}
for an arbitrary assemblage $\mathcal{M}_{(1,2,3)}$. Note that we defined here $\Idiamond^{\mathrm{gen}}(\mathcal{M}) \coloneqq \Ddiamond(\mathcal{M}_{(1,2,3)},\mathcal{M}^{\mathrm{conv}})$ to be the \emph{genuine triplewise incompatibility} of $\mathcal{M}_{(1,2,3)}$, i.e., its distance to the closest assemblage $\mathcal{M}^{\mathrm{conv}} \in \mathrm{JM}^{\mathrm{conv}} \coloneqq \mathrm{Conv}(\mathrm{JM}^{(1,2)},\mathrm{JM}^{(1,3)}, \mathrm{JM}^{(2,3)})$. Furthermore,  $\Idiamond^{\mathrm{pair}}(\mathcal{M}) \coloneqq \Ddiamond(\mathcal{M}^{\mathrm{conv}},\mathcal{M}^{\mathrm{pair}}),$ is the \emph{pairwise incompatibility}, where $\mathcal{M}^{\mathrm{pair}} \in \mathrm{JM}^{pair} \coloneqq \mathrm{JM}^{(1,2)} \cap \mathrm{JM}^{(1,3)} \cap \mathrm{JM}^{(2,3)}$ is the closest assemblage in which all measurements are pairwise-compatible and $\Idiamond^{\mathrm{hol}}(\mathcal{M}) \coloneqq \Idiamond(\mathcal{M}^{\mathrm{pair}})$ is the \emph{hollow incompatibility} of $\mathcal{M}_{(1,2,3)}$. Note that the pairwise and hollow incompatibility depend implicitly on $\mathcal{M}_{(1,2,3)}$.
See also Figure~\textcolor{blue}{$1$} in the main text for the different incompatibility structures. Indeed the incompatibilities defined here, are nothing else but the distances to the next corresponding compatibility structure in Figure~\textcolor{blue}{$1$} in the main text. \\
\indent We now show that the bound in Eq.~\eqref{SupplIncompatiblityDecomposition} is tight for the three Pauli measurements. For simplicity, we focus on the noise-free scenario in the following. From the previous discussions, we know that $\Idiamond(\mathcal{M}_{(1,2,3)}) = \dfrac{1}{2}(1-\eta^*_3)$. \\
\indent For the contribution  $\Idiamond^{\mathrm{gen}}(\mathcal{M}_{(1,2,3)})$ we can use $\mathcal{M}^{\# (1,2)}_{(1,2,3)}$ as (possibily sub-optimal) point in $\mathrm{JM}^{\mathrm{conv}}$. Therefore, we obtain the bound 
\begin{align}
\Idiamond^{\mathrm{gen}}(\mathcal{M}_{(1,2,3)}) \leq \Ddiamond(\mathcal{M}_{(1,2,3)},\mathcal{M}^{\# (1,2)}_{(1,2,3)}) = \tfrac{2}{3}\Idiamond(\mathcal{M}_{(1,2)}) = \tfrac{1}{3}(1-\eta^*_2).    
\end{align} 
\indent For the contribution $\Idiamond^{\mathrm{pair}}(\mathcal{M}_{(1,2,3)})$ we use as a guess for  $\mathcal{M}^{\mathrm{pair}}$ the depolarized version of $\mathcal{M}_{(1,2,3)}$ where it becomes pairwise compatible. That is, $\mathcal{M}^{\mathrm{pair}}$ is of the form
\begin{align}
M^{\mathrm{pair}}_{a \vert x} = \eta^{\mathrm{pair}}_3 \Pi_{a \vert x}+  (1-\eta^{\mathrm{pair}}_3) \dfrac{\mathds{1}}{d}.
\end{align}
Using the results from previous discussions, we therefore obtain 
\begin{align}
\Idiamond^{\mathrm{pair}}(\mathcal{M}_{(1,2,3)}) \leq \dfrac{1}{3}(\eta^*_2-\eta^{\mathrm{pair}}_3)+\dfrac{1}{6}(1-\eta^{\mathrm{pair}}_3),
\end{align}
by bounding  the distance $\Ddiamond(\mathcal{M}^{\# (1,2)}_{(1,2,3)},\mathcal{M}^{\mathrm{pair}})$ through the \ac{SDP} for the diamond norm, i.e., we examine the \ac{SDP} in Eq.~\eqref{Incomp_SDP_primal} with $\mathcal{F} = \mathcal{M}^{\mathrm{pair}}$. \\
\indent Finally, for the contribution $\Idiamond^{\mathrm{hol}}(\mathcal{M}_{(1,2,3)})$ we use as (possibly sub-optimal) candidate for the closest jointly measurable assemblage simply the appropriately depolarized version of $\mathcal{M}_{(1,2,3)}$, i.e., we obtain the bound
$\Idiamond^{\mathrm{hol}}(\mathcal{M}_{(1,2,3)}) \leq \dfrac{1}{2}(\eta^{\mathrm{pair}}_3-\eta^*_3)$. Summing all these bounds up, we obtain that 
\begin{align}
\Idiamond^{\mathrm{gen}}(\mathcal{M}_{(1,2,3)}) + \Idiamond^{\mathrm{pair}}(\mathcal{M}_{(1,2,3)})+\Idiamond^{\mathrm{hol}}(\mathcal{M}_{(1,2,3)}) &\leq \dfrac{1}{3}(1-\eta^*_2) + \dfrac{1}{3}(\eta^*_2-\eta^{\mathrm{pair}}_3)+\dfrac{1}{6}(1-\eta^{\mathrm{pair}}_3)+\dfrac{1}{2}(\eta^{\mathrm{pair}}_3-\eta^*_3) \\
&= \dfrac{1}{2}(1-\eta^*_3), \nonumber
\end{align}
which equals the value for $\Idiamond(\mathcal{M}_{(1,2,3)})$ for the noise free Pauli measurements, as calculated in section \ref{TightnessProof_1}. Therefore Eq.~\eqref{SupplIncompatiblityDecomposition} is tight. Note that the proof crucially relies on knowing the incompatibility of $\Idiamond(\mathcal{M}_{(1,2,3)})$, i.e., without having an analytical expression for this term in higher dimensions $d>2$, this way of proving equality will not work generally. However, we can check numerically, whether the bound in Eq.~\eqref{SupplIncompatiblityDecomposition} is tight for higher dimensional \ac{MUB}. Indeed, our numerics suggest for up to $d=7$ that Eq.~\eqref{SupplIncompatiblityDecomposition} is tight for \ac{MUB} with a deviation of the order $10^{-9}$.

%\includepdf[pages=-]{Paper_draft_distribution_of_incompatibility_PRL_Suppl_post_reports-7.pdf}

\end{document}